\documentclass[nofootinbib,aps,prd,10pt,superscriptaddress,showpacs,preprintnumbers,eqsecnum]{revtex4-1}

\usepackage{amsmath}
\usepackage{dsfont}
\usepackage{multirow}
\usepackage{graphicx}
\usepackage{mathrsfs}
\usepackage{amsfonts}
\usepackage{amssymb}
\usepackage[dvips]{color}

\begin{document}
\newcommand\be{\begin{equation}}
\newcommand\ee{\end{equation}}
\newcommand\bea{\begin{eqnarray}}
\newcommand\eea{\end{eqnarray}}
\newcommand\bseq{\begin{subequations}} %solo con amsmath
\newcommand\eseq{\end{subequations}}
\newcommand\bcas{\begin{cases}}
\newcommand\ecas{\end{cases}}
\newcommand{\p}{\partial}
\newcommand{\f}{\frac}

\title{Periodic orbits in cosmological billiards: the Selberg trace formula for asymptotic Bianchi IX universes, evidence for scars in the wavefunction of the quantum universe and large-scale structure anisotropies of the present universe}

\author{Orchidea Maria Lecian}
\email{lecian@icra.it}
\affiliation{Sapienza University of Rome, Physics Department and ICRA, Piazzale Aldo Moro, 5- 00185 Rome, Italy}

%\today

\begin{abstract}
The Selberg trace formula is specified for cosmological billiards in $4=3+1$ spacetime dimensions. The spectral
 formula is rewritten as an exact sum over the initial conditions for the Einstein field equations
 for which periodic orbits are implied. For this, a suitable density of measure invariant under the billiard
 maps has been defined, within the statistics implied by the BKL (Belinskii, Khalatnikov–Lifshitz) paradigm. 
The trace formula has also been specified for
 the stochastic limit of the dynamics, where the sum over initial conditions has been demonstrated
 to be equivalent to a sum over suitable symmetry operations on the generators of the groups that define 
the billiard dynamics, and acquires different features for the different statistical maps.\\
 A new theoretical interpretation for scars in cosmological billiards is proposed, and numerical evidence is provided.\\
The procedure which links the theoretical framework with the observed large-scale structure of the universe is outlined.\\
 The validity of the Selberg trace formula at the classical level and in
 the quantum regime enforces the validity of the semiclassical descriptions of these systems, thus offering further elements
 for the comparison of quantum-gravity effects and/or the result of compactification of higher-dimensional models with the present observed structure of the universe.\\
This procedure also constitutes
 a new approach in hyperbolic geometry for the application of the Selberg trace formula for a chaotic system whose orbits are 
associated to precise statistical distributions, for both billiard tables corresponding to the desymmetrized fundamental 
domain and to that a a congruence subgroup of it.\end{abstract}

\pacs{ 98.80.Jk Mathematical and relativistic aspects of cosmology- 05.45.-a Nonlinear dynamics and chaos}

\maketitle
\section{Introduction\label{section1}}
%basicbkl
Cosmological billiards arise as the description of the features of space-time in the asymptotic limit towards the cosmological singularity under the BKL (Belinskii, Khalatnikov–Lifshitz) hypothesis, \cite{KB1969a},\cite{Khalatnikov:1969eg},\cite{BK1970},\cite{BLK1971},\cite{Lifshitz:1963ps},\cite{LLK}, for which spacetime points are spatially decoupled within this limit, and the Einstein field equations reduce to s system of ordinary differential equations with respect to time, as time derivatives dominate the dynamic The chaotic motion of a billiard ball in a billiard system, which follows the geodesic evolution of bounces with respect to the (in the limit) infinite potential walls which define the billiard table, is the asymptotic description of the Bianchi IX cosmological model \cite{misn}\cite{Misner:1969hg},\cite{chi1972},\cite{Misner:1994ge} to which the most general anisotropic and homogeneous models are schematized under the BKL paradigm, by the definition of the appropriate statistical maps \cite{Chernoff:1983zz},\cite{isnai83},\cite{isnai85}.\\
The original BKL picture concerns the case of pure gravity in $4=3+1$ space-time dimensions. When also the asymptotic limit of more general inhomogeneous models are dealt with, the appearance of the so-called symmetry walls defines a different (smaller) kind of billiard. For this, one usually refers to the big billiard and the small billiard within all these specifications. The BKL paradigm has proven extremely successful in the description of higher-dimensional systems arising from higher-dimensional unification theories, where new geometrical structures are present, and where a discussion of the physical interpretation of such solutions is based on the proper BKL limit, for which the usual $4=3+1$ dimensional description results as the suitable limit for those physical systems, where a geometry based on the suitable algebraic structures is hypothesized for the target space in which the solution of the Einstein field equations can be represented \cite{Damour:2001sa},\cite{Damour:2000hv},\cite{Damour:2002fz},\cite{Damour:2002cu},\cite{Damour:2002et},\cite{hps2009}. A precise characterization of the $4=3+1$ model descending from these structures has recently been achieved within the framework the the billiard description of the dynamics, for which several symmetry-quotienting mechanisms have been defined according to the geometrical features of the space where the billiards are represented, and according to the Hamiltonian description of the corresponding dynamical systems \cite{Damour:2010sz},\cite{lecianproc}, \cite{Lecian:2013cxa}.\\
%%%%%%%%%%%%%%%%%%%%%%%%%%%%%%%%%%%%%%%%%%%%%%%%%%
The Selberg trace formula \cite{Selberg1956} for billiards on the UPHP \cite{Balazs:1986uj},\cite{Bogomolny:1992cj},\cite{bogomolny1} \cite{venkov1982} is a spectral formula, which allows to write the density of the energy levels of the eigenvalues of the Laplace-Beltrami operator on the UPHP as a sum over the hyperbolic length of the closed geodesics, which constitute the periodic orbits of these billiard systems. The hyperbolic length corresponds to a suitable function of the trace of the composition of matrices which is defined by the suitable iterations of the billiard maps, which, on their turn, describe the periodic trajectory, is invariant under length-preserving transformations in general, and, in particular, under the action of the billiard maps, and under the action of the transformations, which allow for a symmetry-quotienting of the billiard table, when the billiard table does not coincide with the smallest desymmetrized domain of the tessellation of the UPHP. For these purely mathematical features, the Selberg trace formula is valid independently of the physics which is described, i.e. independently of the classical features or of the quantum properties of the system investigated, and, in particular, connects the two different regimes for the semiclassical limit.
%%%%%%%%%%%%%%%%%%%%%%%%%%%%%%%%%
According to these properties, the Selberg trace formula is a very valuable tool for the investigation of the gravitational interaction. Indeed, the schematization of the solution to the Einstein field equations as cosmological billiards allows one to investigate the properties of the energy levels of the periodic orbits for these billiards, the establish to classical features of the motion by the correspondence of the BKL parametrization of periodic orbits as the angular velocity at which the Poincar\'e surface of sections is crossed, at the classical level, to fix these features in the semiclassical limit, and to analyze the possible quantum features of the gravitational interaction below the Planck scale, which corresponds to earliest ages of the universe.\\
For this, the models which predicts quantum-gravity modifications for the spacetime can be linked to the present observation of the actual universe, and phenomenological constraints can be determined for the parameters involved in these models.\\
\\
Form a very different point of view, the analysis of mathematical billiards, i.e. of the billiard system which exhibit chaotic properties, but for which non strong physical characterization is implemented, are usually described, within the quantum regime and the semiclassical limit, by anomalous enhancements of the absolute value of the wavefucntion in correspondence of the lowest periods of periodic orbits. This characterization is usually lost within the classicalized description of the models.\\
\\
The phenomenon of scars \cite{heller} for the wavefunction has been observed in several kinds of billiards, and consists of 'unexpected' enhancements of the absolute value of the wavefunction in correspondence of the lowest-period classical periodic orbits. These phenomenon is understood as an augmentation of the probability for the wavefunction to be located in correspondence of the classical periodic trajectories, even though the maps defining these orbits are unstable, according to the Lyapunov description of the dynamics. The appearance of scars or the wavefunction for the quantum version of billiard systems has been related with the hypothesis that the eigenvalues for the problem associated with the Laplace-Beltrami operator on the UPHP be generated by random-matrix theoretical models \cite{berry} \cite{matrix1} \cite{matrix2}. The specific nature of scars for arithmetical groups has been investigated in \cite{Bogomolny:1992cj} \cite{arith1} \cite{arith2}. The examination of the statistical features of periodic irrationals is still an open project in modern number theories, as the extent of the validity of the Gauss-Kuzmin theorem for this case is still under investigation \cite{mnt}.\\
The phenomenon of scars has also been connected, for cosmological billiards, with particular structures, the 'Farey trees' \cite{cornish}, which can be traced in cosmological billiards when a statistical description different from the BKL prescriptions is considered, as in \cite{cornish} and \cite{lecian}. In \cite{levin2000}, the relation between scars in the wavefunction of the universe and the present large-scale structure of the universe have been outlined, for the case of the compact octagon.\\
The Selberg trace formula allows one to connect the features of the quantum wavefunctions with the mechanisms for which periodic orbits are generated. Within the present work, the appearance of periodic orbits has been connected to the BKL statistics. From the quantum point of view, the presence of scars in correspondence to periodic orbits has also been motivated with different hypotheses \cite{Balazs:1986uj}, \cite{berri}.\\
\\
In \cite{levin2000}, the analysis of scars on a compact octagon has been illustrated to be potentially connected with the present large scale structure of the universe. A characterization of periodic orbits for BKL billiards, on the other hand, has been accomplished by the definition of a Farey map instead of a BKL map, by considering as allowed trajectories only those consisting of purely irrationals, that are the solution to quadratic equations of the oriented endpoints of the corresponding geodesics on the UPHP. A phenomenological connection between the restricted phase space of the BKL dynamics, and that resulting from the first few iterations of the Farey maps has been given in \cite{lecian}.\\\
%%%%%%%%%%%%%%%%%%%%%%%%%%%%%%%%%%%%%%
In the present work, a characterization of the Selberg trace formula for cosmological billiard is given, with respect to the symmetries of the metric tensor. The specification of the Selberg trace formula for mathematical billiards relies on averaging the energy levels of the periodic orbits for the representatives of the conjugacy subclasses of the matrices, whose trace determines the length of the considered closed orbits. Differently form this,  the present analysis is based on specifying the Selberg trace formula for each distinct value of the initial condition to the Einstein field equations, which originate a periodic trajectory, and to rewrite the sum of the spectral formula for the classes of solutions, defined by the analysis of the stochastization of the BKL dynamics. For this, each elements of the representatives of the conjugacy subclasses is given a different weight,a according to the specific features of the BKL dynamics. As a result, numerical evidence is found for scars in the wavefucntion of the universe. The theoretical interpretations of these scars finds agreement with the conjecture of \cite{berri}, for which scars in the quantum models and in the semiclassical limit are found for the trajectories of the phase space, where the classical system happens to 'spend the longest time'.
\\
The invariance of the WDW equation under the iterations of the billiard maps defining periodic trajectories enforces the physical interpretation of the semiclassical limit and allows one to directly relate the classical periodic trajectories with the phenomenon of scars in the wavefunction observed in the quantum regime.\\
\\
The semiclassical limit for the wavefunction of the universe is relevant in describing the features of the gravitational interaction for the transition from the quantum regime to the classical phase. At these level, the emergence of quantum-gravity effects id related to the modifications of the underlying space-time geometry, and term of the Selberg trace formula accounting for quantum effects can be isolated and considered as a modification of the spacing of the energy spectrum due to these geometrical effects.\\
Indeed, the semiclassical limit of the wavefunction of the universe obtained form the WDW equation is usually considered for the verification of quantum-gravity effects.\\
\\
In the quantum regime, the Selberg trace formula has two key roles.\\
On the one hand, the definition of periodic trajectories for the quantum regime allows one to interpret the quantum maps for the energy levels which are considered. The definition of quantum BKL maps for the wavefunction then relates the corresponding observables, defined by the squared module of the wavefunction evaluated under the pertinent regions of the UPHP or of the restricted phase space, with a precise energy level of the energy spectrum. Quantum BKL numbers are then defined as the angular velocity at which the billiard ball crosses the Poincar\'e surface of section, for which the Poincar\'e return map is implemented, and, as a consequence of the properties of the hyperbolic length of a trajectory of being independent of the physical regime at which the trajectory is considered, as well as a consequence of the features of the Selberg trace formula, for which the eigenvalues of the Laplace-Beltrami operator define such an angular velocity, the definition of quantum BKL numbers is well posed, and the physical meaning of trajectories under the BKL map is valid both at classical level and at quantum regime.\\
The presence of strong quantum-gravity effects  \cite{amati}, \cite{8}, \cite{ash2}, \cite{ash1} has been described as deeply modifying the chaotic dynamics characterizing the cosmological singularity. The analysis of the spectral distribution of the energy level defines a quantitative comparison for the phenomena, which are supposed to modify  the cosmological singularity.\\
\\
The general cosmological solution, described under the BKL paradigm, needs to be related with the complete thermal history of the universe, and to be connected with the results of present observations. For this, a suitable quasi-isotropization mechanism has to be hypothesized, such that the strongly anisotropic oscillating regime is reconducted to a non-oscillating behavior for the scale factors. In the present work, a procedure is outlined, for which these connections can be found. Indeed, the anisotropic Sky patterns described in observational evidence can provide with the items of information which characterize  both a suitable range of values for the three BKL direction are specified at the time when the external contributions to the Einstein field equations are considered, and the level of stochastization of the BKL dynamics. The geometry of the target space for the parametrization of the solution to the Einstein field equations allows one to evaluate the age of the universe at which the external contributions have started playing a significant role in 'freezing' the features of the oscillating anisotropic BKL regime by the (hyperbolic) length of the geodesics path followed by the billiard ball.\\
According to the mathematical features of the UPHP encoded in the Selberg trace formula, it is straightforward to recognize that the BKL probabilities defined for the one-variable map and for the two-variable map, normalized according a given symmetry operation on the generators of the billiard maps, defined in \cite{leciannew}, are the two-point correlation function for the most general anisotropic cosmologies: the BKL probabilities for the big billiard are the two-point correlation function for anisotropic and homogeneous cosmologies, while the BKL probabilities for the small billiard are the two-point correlation function for anisotropic inhomogeneous cosmologies. The choice whether to take into account the variable $u^-$ corresponds to a means to define the degree of stochasticity that has to be described, and has to be physically characterized as an intrinsic parameter describing the features of the spacetime in the vicinity of the cosmological singularity, which cannot be modified within the evolution of the dynamics. Differently, the several steps for the definition of the stochastic limits are means to master the 'speed' at which the completely stochastic regime is eventually reached. The suitable combination of all these theoretical characterizations of the two-point correlation functions for anisotropic cosmologies allows one to very precisely implement a comparison with the observed anisotropies in the present Sky by means of a comparison with the two-point correlation function evaluated for particular effects of the gravitational interaction in the large-scale structure of the universe, and to establish the age of the universe, at which a quasi-isotropization mechanism has commenced determining the nowadays anisotropy \cite{9a}, \cite{9b}, \cite{9c}, \cite{Shafieloo:2011sd}.\\
\\
The paper is organized as follows.\\
In Section \ref{section2}, the description of the features of the universe in the vicinity of the cosmological singularity within the framework of the BKL paradigm are recalled, and particular attention is devoted to cosmological billiards.\\
In Section \ref{section3}, the main features of the Selberg trace formula for billiard systems are revised, and the particular assumption of the description of billiards as with a completely stochastized Markovian dynamics is reported.\\
In Section \ref{section4}, the Selberg trace formula for cosmological billiards is written as a sum over the initial conditions of the Einstein field equations, and the corresponding expression is shown to be equivalently stated by the a sum over the BKL probabilities for the corresponding geodesics to take place, as described by the BKL statistics. The spectral formula has also been rewritten for the limit of a completely stochastized BKL process.\\
In Section \ref{section5}, numerical evidence for the phenomenon of scars in cosmological billiards is presented. A new theoretical explanation for this phenomenon is provided by the physical implications of the Selberg trace formula for cosmological billiards, where the physical characterization of the trajectories defines the relative probabilities between the different geodesic paths, which have been considered equivalent in previous literature. The role of the stochastization of the BKL dynamics in the scars of the wavefucntion of the universe is analyzed.\\
In Section \ref{section6}, the connection between scars in the wavefucntion of the universe, the physical qualification of the billiard trajectories within the Selberg trace formula and the anisotropic sky patterns provided by observational evidence have been connected by outlining the precise procedure which allows one to establish the age of the universe at which a quasi-isotropization mechanism has started modifying the original BKL dynamics, and the corresponding degree of stochastization, allowing for the definition of the origin of the observed anisotropic pattern, by discriminating between quantum effects or classical effects.
\section{Cosmological billiards\label{section2}}
Cosmological billiards are a schematization of the asymptotic limit toward the cosmological singularity of the solution to the Einstein field equations under the hypothesis of the BKL paradigm, for which spacetime points are spatially decoupled in the vicinity of the cosmological singularity. In particular, the solution to the Einstein field equations can be mapped to a suitable target space, endowed with Lorentzian metric, where the parametrization of the symmetries of the solution to the Einstein field equations corresponds to the geodesic motion of a billiard ball within a unit hyperboloid (or hyper-hyperboloid if a higher number of space dimensions is considered). The motion inside this unit hyperboloid, characterized by bounces on the surface, is in one-to-one correspondence with this limit of the solution to the Einstein field equations, which write, for each approximation to a diagonal form of the space part of the metric tensor $g_{ij}\equiv diag (a, b, c)$ (in the suitable time gauge) as
\begin{subequations}\label{abc}
\begin{align}
&2\frac{d^2\ln a}{d\tau^2}=(b^2-c^2)^2-a^4,\\
&2\frac{d^2\ln b}{d\tau^2}=(c^2-a^2)^2-b^4,\\
&2\frac{d^2\ln c}{d\tau^2}=(a^2-b^2)^2-c^4,
\end{align}
\end{subequations}
such that each of the approximations can be parameterized by the statistical variable $u$, which corresponds to the variable $u^+$ of the restricted phase space of the model (defined in the following), and to which a supplementary statistical variable $u^-$, which has its own geometrical interpretation in the restricted phase space, and which has the physical role of keeping track of the evolution of the billiard system, and the statistical role of allowing the analysis of the stationary limit of the iterations of a statistical map on the set of the two variable $u^+$ and $u^-$.\\
The solution of the Hamiltonian constraint for these models corresponds to projecting the motion of the billiard ball onto the unit hyperboloid; the restricted phase space is obtained by considering a fixed energy shell for the system (from the solution to the Hamiltonian constraint, for the morphology of the Einstein field equations), and to the choice of a particular Poincar\'e surface of section, for which the continuous chaotic motion can be schematized according to the Poincar\'e return map, characterized by the variables $u^+$ and $u^-$, of the billiard ball through this cross section.\\
Considering different potential terms in the gravitational action corresponds to obtaining different asymptotic limits toward the cosmological singularity of the shape of the potential walls, which enclose the motion of the billiard ball for the motion within the unit hyperboloid as well as for the projected version on the hyperboloid, and by means of the suitable geometrical transformations, which do not modify any aspects of the dynamics, but allow for a more comfortable visualization of the system, on the Kasner circle and then on the Upper Poincar\'e Half Plane.\\
The precise establishment of the big billiard, which corresponds to the pure gravitational case as far as the specification of the action for the gravitational filed is concerned, and of the small billiard, for which also centrifugal terms are considered in the gravitational action, descends directly form the most general characterization of cosmological billiards. A characterization of the billiard maps on the UPHP is obtained in \cite{Lecian:2013cxa}, while an analysis of the problems related with the definition of a Poincar\'e surface of section is expressed in \cite{lecian}.
\subsection{The broad panorama of cosmological billiards}
A description of the Kasner circle, which is, to a precise extent, as far as the dynamical properties of cosmological billiards are concerned, dual to the present one with respect to the physical interpretation of the properties of the Kasner circle, is due to \cite{wain} \cite{uggla2003},\cite{ringstrom2000}, \cite{Ringstrom:2000mk}, \cite{Heinzle:2009eh} \cite{uggla2012} \cite{wct}, \cite{uggla2013}, \cite{coley}
within this latter framework, the appearance of spikes, i.e. anomalous 'jumps' of the statistical variables, can be encoded within the properties of certain solution-generating techniques, to be effective in phenomenologically modifying the structure constants of the Bianchi classification of the spacetime, according to precise patterns. The corresponding Petrov classification has been proposed in \cite{Bini:2008qg},
\cite{Cherubini:2004yi}. Numerical simulation of these anomalous behavior reveals these properties. Numerical investigation of the properties of the generic cosmological solution also reveals particular features of possible statistical parametrization of the solution to the Einstein file equations
 \cite{garf93}, \cite{berger1993} \cite{berger1994}, \cite{berger1997}.\\
A physical characterization of the algebraic structures of  cosmological billiards has been accomplished by \cite{Fleig:2011mu}, \cite{carb}, \cite{berger1991} \cite{fre1},\cite{fre2}.\\
The quantum features implied by the solution of the WDW equations, as well as the mathematical properties of the WDW equation have been investigated in , while the mathematical features of the WDW equations have been defined in \cite{primordial}, \cite{puzio}, \cite{monix}, the mathematical features of the WDW equation for cosmological billiards have been analyzed in \cite{Kleinschmidt:2009cv}, \cite{mk11}, and the interpretative problems brought by the definition of a wavefucntion of the universe have been uncovered in  \cite{gibb}, \cite{Graham:1990jd}, \cite{isham1} \cite{isham}.\\
The features of the wavefucntion of a universe characterized by a low degree of anisotropy has been investigated in \cite{hawking84}, \cite{amsterdamski85}, \cite{moss85}, \cite{furusawa86}, \cite{furusawa861}; in \cite{berger}, the features of wavefucntions corresponding to long eras have been clarified to imply a fragmentation of the wavefucntions for this statistical characterization. Boundary conditions for the wavefucntion of the universe have been discussed and compared in \cite{csordas1991}, \cite{Kleinschmidt:2009hv}, \cite{Forte:2008jr}, \cite{Graham:1990jd}, \cite{graham1991},\cite{Kleinschmidt:2010bk}, and in \cite{Benini:2006xu} a wavefunction on a distorted domain is considered.\\
The BKL dynamics find a graceful description also in terms of the Misner variables and of the Misner-Chitr\'e variables in $4=3+1$ spacetime dimensions, by means of which the general validity of the BKL paradigm in the definition of the asymptotic silence have been tested also by hypothesisizng a different characterization for the killing fields that define the $3$ geometry of the metric tensor.\\ 
\subsection{The big billiard}
\begin{figure*}[htbp]
\begin{center}
 \includegraphics[width=0.7\textwidth]{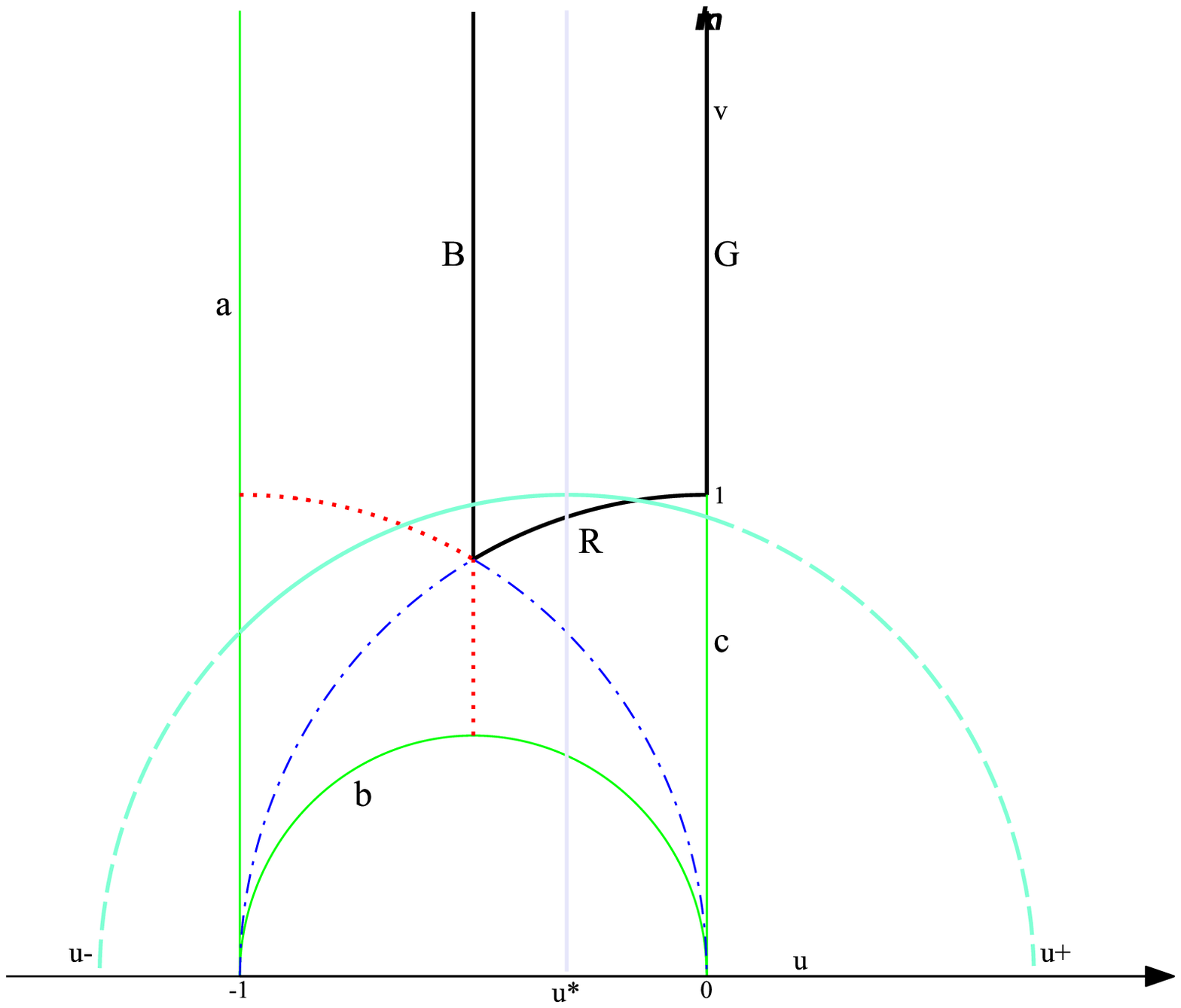}
\caption{\label{figura1} The billiard tables on the UPHP. The big billiard table is delimited by the sides $a$, $b$ and $c$; the subdominant symmetry walls consist of the blue (dashdot) lines bisecting the corners of the big billiard, and of the red (dotted) lines perpendicular to the sides of the billiard, while the dominant symmetry walls define the small billiard table, delimited by the sides $R$, $G$ and $B$. An epoch of the $ba$ type is sketched (orange circle), and is parametrized by the oriented endpoints $u^+$ and $u^-$ of the corresponding geodesics (dashed circle). A generic Poincar\'e surface of section $u^*$, here a generalized geodesics, is represented by the violet ('vertical') line.}
\end{center}
\end{figure*}

The big-billiard is delimited by the three infinite potential walls $a$, $b$, $c$, illustrated in Figure \ref{figura1},  
\begin{subequations}\label{bbu}
\begin{align}
&a: u=0\\
&b: u=-1\\
&c: u^2+u+v^2=0, 
\end{align}
\end{subequations}
for which define the reflections for the variable $z=u+iv$ coordinatizing the UPHP as
\begin{subequations}\label{bbz}
\begin{align}
&Az=-\bar{z},\\
&Bz=-\bar{z}-2,\\
&Cz=-\tfrac{\bar{z}}{2\bar{z}+1},
\end{align}
\end{subequations}
i.e. unquotiented big-billiard map $\mathcal{T}$.\\
\\
The quotiented big-billiard map for $z$ reads
\begin{equation}\label{BKLz}
z\rightarrow T^{-1}z\rightarrow ... \rightarrow T^{-n+1}z\rightarrow z'\equiv\frac{1}{\bar{z}-n+1}-1\equiv T^{-1}SR_1T^{-n+1}z,
\end{equation}
where the first part of (\ref{BKLz}) is the Kasner quotiented BKL epoch map, while the second part of (\ref{BKLz}) is the Kasner quotiented CB-LKSKS map.\\ 
The unquotiented big- billiard map and the quotiented big-billiard one-variable map for the variable $u^+$, i.e. 
\begin{equation}\label{BKLup}
u^+\rightarrow u^+-1\rightarrow ... \rightarrow u^+-(n-1)\rightarrow u^{+ '}\equiv\frac{1}{u^+-n+1}-1.
\end{equation}
and the quotiented big-billiard two-variable map for the variables $u^+$ and $u^-$, i.e.
\begin{equation}\label{BKLupm}
u^\pm\rightarrow u^\pm-1\rightarrow ... \rightarrow u^\pm-(n-1)\rightarrow u^{\pm '}\equiv\frac{1}{u^\pm-n+1}-1,
\end{equation}
are recast by setting $v=0$ in (\ref{BKLz}), such that the results of \cite{Damour:2010sz} and \cite{Lecian:2013cxa} are reconstructed. These two maps act on the region of the restricted phase space illustrated in Figure \ref{figura2}.\\
\\
The BKL probability for an era to contain $n$ epochs reads
\be\label{probbb}
P^{BKL}(n)=\tfrac{1}{\ln2}\int_{-2}^{-1}du^-\int_{n-1}^n\tfrac{du^+}{(u^+-u^-)^2}=\tfrac{1}{\ln 2}\ln\tfrac{(n+1)^2}{n(n+2)},
\ee 
while the BKL probability for a sequence of eras $k$ containing $n_1, n_2, ..., n_k$ epochs each writes as the integral on the corresponding subregions of the restricted phase space of the invariant density of measure for the BKL map for the big billiard $W^{u^+}$ for the variable $u^+$ as
\be\label{probb}
P^{BKL}(n_1, n_2, ..., n_k)=\tfrac{1}{\ln2}\int_{-2}^{-1}du^-\int_{n_1-1+\tfrac{1}{n_2+\tfrac{...}{n_k-1}}}^{n_1-1+\tfrac{1}{n_2+\tfrac{...}{n_k}}}\tfrac{du^+}{(u^+-u^-)^2}.
\ee
The two-variable map (\ref{BKLupm}) allows \cite{leciannew} for a comparison of different periodic trajectories defined by a repetition of the sequence $k$ according to a different number of iterates of the two-variable map by means of the definition of the BKL probabilities normalized according to the statistics of (\ref{BKLupm}) as 
\be\label{probcyc}
P^{BKL}_{2-var}(k)\equiv P^{BKL}_{cycl}(k)\equiv\tfrac{P^{BKL}(n_1, n_2, ..., n_k)}{\sum_{cycl(k)}P^{BKL}(n_1, n_2, ..., n_k)},
\ee
where, in the denominator, a normalization according to the cyclic permutations of the elements of the sequence $k$ is implied. Different non-cyclic permutations on the elements of the periodic sequence are not compatible with the two-variable map in the definition of these probabilities \cite{leciannew}.\\
Differently, the one-variable map (\ref{BKLup}) allows for a comparison of different trajectories defined by a 'reshuffling' of the elements of the sequence $k$, where the consistency of this operation is ensures by the marginalization of the statistical variable $u^-$ with respect to the previous case, as the variable $u^-$ encodes the 'memory' of the billiard system about the 'past' evolution of the dynamics. For this, BKL probabilities for the one-variable map (\ref{BKLup}) are defined as as 
\be\label{probper}
P^{BKL}_{1-var}(k)\equiv P^{BKL}_{per}(k)\equiv\tfrac{P^{BKL}(n_1, n_2, ..., n_k)}{\sum_{per(k)}P^{BKL}(n_1, n_2, ..., n_k)},
\ee
where, in the denominator, a normalization according to all the permutations of the elements of the sequence $k$ is implied.\\
%%%%%%%%%%%%%%%%%%%%%%%%%%%%%%%%%%%%%%%%%%%%%%%%%%%%%%%%%%%%
\subsection{The small billiard}
\begin{figure*}[htbp]
\begin{center}
\includegraphics[width=0.7\textwidth]{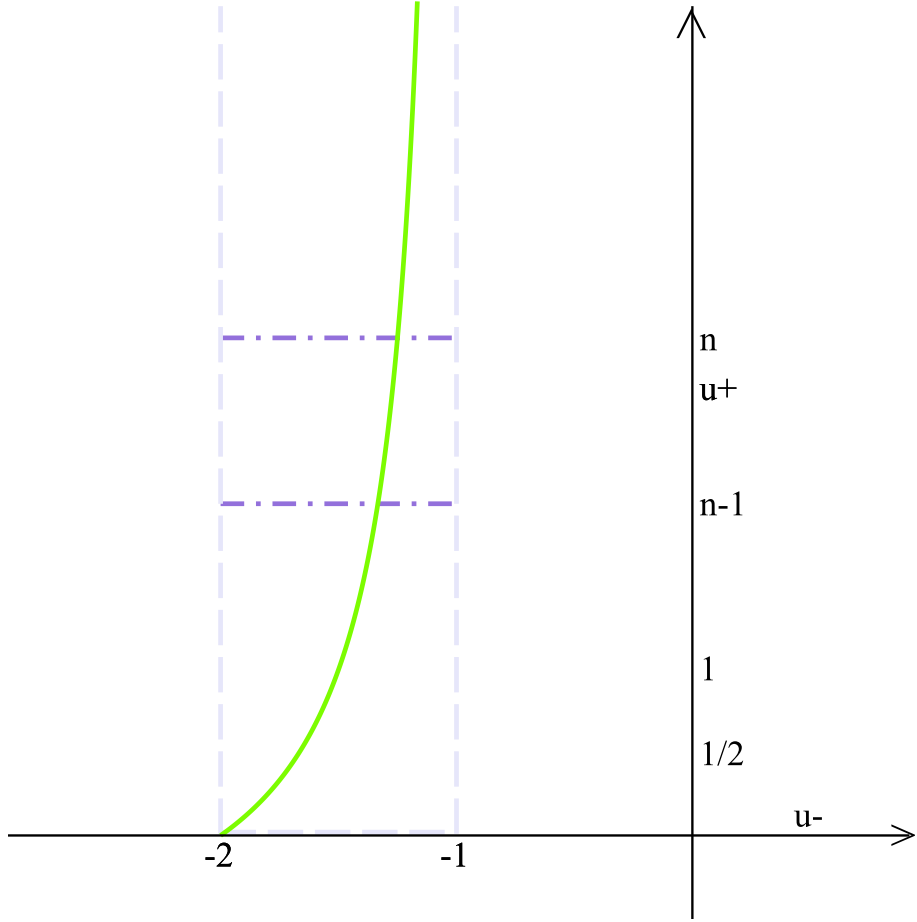}
\caption{\label{figura2} The restricted phase space for cosmological billiards is parameterized by the variables $u^+$ and $u^-$. The regions available for the quotiented dynamics are illustrated. The subregion corresponding to the Kasner quotiented CB-LKSKS era map for the big billiard is delimited by the dashed lines. This region is divided by the function $u_\gamma$ (plotted by the solid line) into two different subregions, defined in (\ref{regsb}), which correspond to the Kasner projection of the dynamical subregions of the restricted phase, where the small billiard map is defined, on the UPHP, by a different number of Weyl reflections for the length of the corresponding era in the big billiard. The region $I$ of (\ref{regsb}) is the subdomain enclosed between $u_\gamma$ and $u^+=0$, while the region $II$ is enclosed between $u^+=\infty$ and $u_\gamma$. The dashdot lines enclose the generic sub-box, for which the probability to find an era containing $n$ epochs is defined, both for the big billiard, and for the small billiard, for the BKL probabilities for the big billiard $P^{BKL}(n)$ (\ref{probb}), and for the BKL probabilities for the small billiard ${\rm p}^{BKL}(n)$ in (\ref{smallprob}), respectively.}
\end{center}
\end{figure*}

The small billiard table is defined by the sides $G$, $B$, $R$, defined as 
\begin{subequations}\label{smdomain}
\begin{align}
&G: \ \ u=0,\\
&B: \ \ u=-\tfrac{1}{2},\\
&R: \ \ u^2+v^2=1,
\end{align}
\end{subequations}
and illustrated in Figure \ref{figura1}, for which the following reflections are defined on the variable $z$
\begin{subequations}\label{smalltrasf}
\begin{align}
&R_1(z)=-\bar{z},\\
&R_2(z)=-\bar{z}+1,\\
&R_3(z)=\tfrac{1}{\bar{z}},
\end{align}
\end{subequations}
which are usually named as the small-billiard map on the UPHP: According to the most general classification \cite{Kleinschmidt:2009cv}, $R$ are Weyl reflections, and $B$ and $G$ are affine reflections.\\
The CB-LKSKS map for the small billiard, $t_{CB-LKSKS}$, is defined by two different kinds of transformations \cite{leciannew}, denoted by $\mu=I, II$, i.e., 
\begin{subequations}\label{tsb}
\begin{align}
&t^{I}z=T^{-1}SR_1T^{-n+1}z,\ \ {\rm for} (u^+, u^-)\in S^{1}_{ba} {\rm and} (u^+, u^-)\in S^{2}_{ba},\label{tsb1}\\
&t^{II}z=T^{-1}SR_1T^{-n+1}R_3z,\ \ {\rm for} (u^+, u^-)\in S^{2'}_{ba}, (u^+, u^-)\in S^{3}_{ba}, {\rm and} (u^+, u^-)\in S^{3'}_{ba}\label{tsb2}.
\end{align}
\end{subequations}
which act on the subregions $\mu=I, II$ of the reduced phase space
\begin{subequations}\label{regsb}
\begin{align}
&I: -2\le u^-\le-1,\ \ u_\gamma\le u^+\le \infty,\\
&II: -2\le u^-\le-1,\ \ 0 u^+\le u_\gamma,
\end{align}
\end{subequations} 
as illustrated in Figure \ref{figura2}, where the function $u_\gamma$ is defined as
\be
u_\gamma(u^+):\ \ u^+=-\tfrac{u^-+2}{u^-+1},
\ee 
such that the values of $\mu$ correspond to the number of Weyl reflections $R$ in the quotiented small-billiard map for the variable $z$.\\
The BKL probabilities ${\rm p}^{BKL}_\mu(n)$ for an $n$-epoch era to occur are expressed as 
\begin{subequations}\label{smallprob}
\begin{align}
&{\rm p}^{BKL}_{I}(n)=\tfrac{1}{2\ln 2}\ln\tfrac{(n+1)^2(n^2+1)}{n^2(n^2+2n+2)}\\
&{\rm p}^{BKL}_{II}(n)=\tfrac{1}{2\ln 2}\ln\tfrac{(n+1)^2(n^2+2n+2)}{n^2(n^2+1)(n+2)^2},
\end{align}
\end{subequations}
while the BKL probabilities ${\rm p}^{BKL}_\mu(k)$ for a sequence of eras containing the sequence $k$ of epochs  to occur are expressed as the integral of the pertinent density of invariant measure $W^{\mu}(u^+)$ for the variable $u^+$ on the corresponding subregions of the restricted phase space,
\begin{subequations}\label{smallprobsequence}
\begin{align}
&{\rm p}^{BKL}_{I}(n_1, n_2, ..., n_k)\equiv\int_{\tfrac{1}{n_2+\tfrac{...}{n_k-1}}}^{\tfrac{1}{n_2+\tfrac{...}{n_k}}}W^{I}( u^+)du^+,\\
&{\rm p}^{BKL}_{II}(n_1, n_2, ..., n_k)\equiv\int_{\tfrac{1}{n_2+\tfrac{...}{n_k-1}}}^{\tfrac{1}{n_2+\tfrac{...}{n_k}}}W^{II}(u^+)du^+,
\end{align}
\end{subequations}
as illustrated in Figure \ref{figura2}. The detailed expressions for the densities of invariant measures $W(u+)$ and of the stochastic features of the BKL dynamics is given in \cite{leciannew}, as well as the definitions for the BKL probabilities for the different symmetry operations
\be
{\rm p}^{BKL}_{\mu 2-var}(k)\equiv {\rm p}^{BKL}_{\mu cycl}(k)\equiv\tfrac{{\rm p}^{BKL}_\mu(n_1, n_2, ..., n_k)}{\sum_{cycl(k)}{\rm p}^{BKL}_\mu(n_1, n_2, ..., n_k)},
\ee
and 
\be
{\rm p}^{BKL}_{\mu 1-var}(k)\equiv {\rm p}^{BKL}_{\mu per}(k)\equiv\tfrac{{\rm p}^{BKL}_\mu(n_1, n_2, ..., n_k)}{\sum_{per(k)}{\rm p}^{BKL}_\mu(n_1, n_2, ..., n_k)}.
\ee

\subsection{Quantum regime}
The quantum regime of cosmological billiards is defined within the framework of the analysis of the effects of a quantum version of the space-time in the vicinity of the cosmological singularity on the features of the quantum wavefunction for the universe, as described by the WDW equation, and on the implications that the quantum description has o the properties of this asymptotic limit of the Einstein field equations.\\
The full quantum regime of the model is accounted for by a description of the corresponding eigenvalue problem for the variables $\rho$ and $\gamma$, which describe the radial part and the angular part of the parametrization of the target space, in which the chaotic description is represented.\\
The solution of the Hamiltonian constraint by the elimination of the radial variable from the mathematical investigation, i.e. by the definition of a suitable function of $\rho$ for which the Hamiltonian constraint an be solved by projecting the motion onto the surface of the unit hyperboloid, on its turn allow one to eliminate the variable $\rho$ from the WDW equation. The well-behaviored-ness of the wavefunction (factorized for the two variables) REF ensures that the projection on the Unit Hyperboloid be fully consistent \cite{Kleinschmidt:2009cv} \cite{mk11}.\\
\\
The WDW equation for cosmological billiards on the UPHP rewrites as the eigenvalue problem associated with the Laplace- Beltrami operator on these geometries. The analysis of the geometrical features of these spaces allows one select from them the features relevant n the description of the dynamics. As a result, the symmetries of the wavefunction for the big billiard and for the small billiard in the symmetry-quotiented versions of the dynamics reveal that such wavefunctions are invariant under a class of transformations which is smaller with respect to those that characterize the UPHP. The symmetries of the wavefunction of the universe must therefore be the same as those conserved in this procedure. These symmetries express in the quantum regime the symmetries of the metric tensor as stated by the solution of the Einstein field equations, and are taken care of by the statistical maps once the three asymptotic projective anisotropy directions have been fixed by the picture of the walls delimiting the billiard table.\\
As a result,
the WDW equation reads
\begin{equation}\label{eigen}
-\Delta_{\rm LB}\Psi(z)=E\Psi(z)
\end{equation}
where the wavefunction is decomposed according to
$\Psi=\sum_s \Phi_s$
\begin{equation}\label{maass}
\Phi_s(u,v; c_1, c_2)=c_1v^s+c_2v^{1-s}+\sum_\mu c_\mu y^{1/2}\mathcal{K}_{s-\tfrac{1}{2}}(2\pi\mid \mu\mid v)exp[2\pi i\mu u],
\end{equation}
, and the wavefunction writes
The eigenvalue problem (\ref{eigen}) for the Fourier-expanded Maass waveforms (\ref{maass}) is then
\begin{equation}
-\Delta_{\rm LB}\Phi_s(u,v)=E_s\Psi_s(u,v),
\end{equation} 
where the eigenvalues $E_s$ define the energy spectrum of the quantum Hamiltonian, $E_s\equiv s(s-1)$. In (\ref{maass}), a Fourier decomposition is implied for the variable $u$, and describes the properties of classical trajectories as (generalized) circles, while the symmetries of the $v$ direction are encoded in the decomposition of modified Bessel functions of the second kind $\mathcal{K}$, which, on their turn, describe the topology of the full unprojected space of the logarithmic scale factors parameterized by he variable $\rho$ and $\gamma$. In other words, the behavior of the wavefunctions with respect to the variable $v$ is one implied by projecting out the variable $\rho$ form the schematization of the motion in the solution of th e Hamiltonian constraint.\\
%%%%%%%%%%%%%%%%%%%%%%%%%%%%%%
\subsection{Periodicity phenomena for the unquotiented big billiard\label{machinery}}
Periodic orbits of the big billiard are a phenomenon which is more complicated than its symmetry-quotiented versions. Given a $m$-periodic orbit of the one-dimensional BKL epoch map
\begin{equation}\label{eq32}
T_{\rm BKL}^m (u^+) = u^+ 
\end{equation}
with 
\begin{equation}\label{eq33}
\sum_{i=1}^{1=k}n_i=m,
\end{equation}
periodic orbits of the big billiard group are given by $mp$ iterations of the unquotiented billiard map $\mathcal{T}$
\begin{equation}\label{mp}
\mathcal{T}^{mp} u=u  ,
\end{equation}
where $m$ is the total number of BKL epochs for which the quotiented big billiard BKL map is periodic, and $p$ is the order of the Kasner transformation for which the new sequence of eras takes place in the correct corner and in the correct orientation.\\
\\ 
According to the unquotiented big billiard described by Eq.'s (\ref{bbz}), the iteration of the unquotiented big billiard map (\ref{mp}) contains an even number of reflections (whose composition results as an absence of reflections) if the product $mp$ is even, i.e. for $mp=2j$ for some $j\in \mathbb{N}$, while this iteration contains an odd number of Weyl reflections (whose composition results as the presence of a Weyl reflection) if the product $mp$ is odd, i.e. for $mp=2j+1$ for some $j\in \mathbb{N}$.\\
\\
The exact sequence of transformations that compose the unquotiented big billiard map ${\mathcal T}^{mp}$ is reconstructed by evaluating the rotation induced by any new era, and its relative orientation with respect to the previous one, by means of the knowledge of the exact sequence of the $n_i$.\\
A direct implication of this method is the following. Let $M$ be the number of eras containing an even number of epochs, with $M\le m$. Then one learns that $p=2$ iff $M$ is odd, while $p=1$ or $p=3$ if $M$ is even. Furthermore, given $M$ even, $p=1$ for $m=0$ mod $3$, while $p=3$ for $m=1,2$ mod $3$.
%%%%%%%%%%%%%%%%%%%%%%%%%%%%%%%%%%%%%%%%%%%%%%%%%%5
\subsection{Boundary conditions}
Evan thought the definition of the Selberg trace formula is independent of the regime at which the dynamics is considered from a physical point of view, i.e. it holds in the classical billiard description, as well as in the quantum version of the dynamics and in the semiclassical limit, suitable boundary conditions have to be imposed for the quantum wavefunction, when the quantum regime is envisaged.\\
The validity of the definition of a quantum wavefucntion for cosmological billiards has been analyzed to be independent of the boundary conditions one choses to impose on it, as far as the physical characterization of BKL quantum numbers. In fact, these quantum numbers are defined according to their capability in encoding, at the semiclassical limit, the different angular velocity at which a classical trajectory crosses the Poincar\'e surface of section, for which the discrete map is obtained, as this cross section can be chosen as different from the sides of the billiard table, such that the quantum regime is implemented according to this semiclassical interpretation. Furthermore, the Selberg trace formula can be shown to be defined, in the semiclassical limit, through the proper term of the WKB approximations, as equal to that obtained in the quantum model after a complete calculation is performed. Nevertheless, the choice of boundary conditions not only  implies the wavefucntions to acquire different values on the boundaries of the billiard table, but also is, in principle, able to modify the definition of the quantities contained in the Selberg trace formula.\\
The choices of Dirichlet boundary conditions \cite{csordas1991}, \cite{Benini:2006xu} \cite{Kleinschmidt:2009hv} or Neumann boundary conditions \cite{Forte:2008jr} \cite{Graham:1990jd} boundary conditions have been critically compared in \cite{graham1991} and discussed \cite{Kleinschmidt:2010bk}.\\
A discussion of the properties of the boundary conditions, in the quantum regime and in the semiclassical limit, and a comparison between the symmetries of the boundaries of the billiard tables and the symmetries of the wavefunctions, which must respect the symmetries of the classical dynamics and match the quantum one in the semiclassical limit, is performed in \cite{Lecian:2013cxa} for the particular cases of the small billiard and of the big billiard.\\
%%%%%%%%%%%%%%%%%%%%%%%%%%%%%%%%%%%%
The analysis of the previous sections allow one to specify the Selberg trace formula for cosmological billiards.\\
The very particular features of cosmological billiards allow one to write this trace formula according to different aspects of the dynamics, which are unexplored in the previous studies on 'mathematical billiards', thus allowing one to connect the geometrical properties of the spectrum of the eigenvalues of the Laplace-Beltrami operator to the statistical properties of the billiard maps, which reproduce the symmetries of the solution to the Einstein field equations.
\section{The Selberg trace formula\label{section3}}
The Selberg trace formula is based on the sum over (a suitable function of) the hyperbolic length of the closed geodesics corresponding to one period of a periodic orbit, which, on its turn, represents the periodic sequence on which the initial value of the variable $u^+$ is defined.\\
As recalled in the previous Section, the properties of the spacing of the eigenvalues of the Laplace-Beltrami operator on the UPHP depend on the composition of matrices which define the periodicity condition and not on the initial values of the oriented endpoints of the geodesics, which define the periodic orbit. This way, all the quantity needed in the Selberg trace formula, i.e. the prefactors accounting for the number of reflections and the hyperbolic length of the closed geodesics, can be stated for any property of the dynamics, as it will be clarified in the following. As a result, the Selberg trace formula can be considered as a method to classify the initial conditions for the Einstein field equations in the asymptotic limit to the cosmological singularity, and to relate them with the composition of matrices which define the periodic sequences.\\
The key step in the definition of the Selberg trace formula for cosmological billiards is the understanding the a sum over (a suitable function of) the hyperbolic lengths of periodic orbits is related with the definition of the finite sequence of eras $k$ defining each closed orbit int the periodic sequence, for which the periodic value of the continued fraction decomposition of $u^+$ is defined.\\
\\
%%%%%%%%%%%%%%%%%%%%%%%%%%%%%%%
The asymptotic (towards the singularity) limit of the Bianchi IX universes, both in the homogeneous case and in the inhomogeneous one, constitute very particular models, for which the Selberg trace formula can be computed exactly.\\
Indeed, the physical characterization of the trajectories, as well as the symmetry-quotienting mechanisms which have been implemented to outline particular information about the behavior of periodic orbits, allow one to rewrite the Selberg trace formula exactly, i.e. by connecting the hyperbolic length of the closed trajectories and their (normalized) probability without averaging on a large number of succession of eras.\\
The Selberg trace formula is a spectral formula which defines the spacing between the eigenvalues of the Laplace-Beltrami operator by (a suitable function of) the hyperbolic length of periodic orbits.
The trace of a (composition of) matrices defining a periodic orbit is invariant under symmetry-preserving transformations. For each such matrices $\mathcal{Q}$, the hyperbolic length of a closed orbit $L$ is connected to the trace $\mathcal{M}$ of $\mathcal{Q}$ as
\begin{equation}\label{eq8}
 2cosh \tfrac{L}{2}=\mid Tr \mathcal{Q} \mid \equiv \mathcal{M }.
\end{equation}
\\
According only to this specification, the spectrum of the eigenvalues of the Laplace-Beltrami operator are given by the expression 
\begin{equation}\label{stf}
 d(E)=\langle d(E)\rangle+\tilde{d}(E)+d_{osc}(E),
\end{equation}
where, for each energy level, $d(E)=\sum_{n}\delta(E-E_n)$.\\
\\
Given $g(l)$ the number of periodic orbits of length $l$, and given $g(n)$ the number of distinct conjugacy classes corresponding to a sequence of matrices with trace $n$, as in (\ref{eq8}), in the limit $n\rightarrow\infty$, the mean multiplicity of periodic orbits is
\begin{equation}\label{eq9}
 \langle g(n)\rangle =\tfrac{n}{\ln n}
\end{equation}
This way, the normalized number of distinct conjugacy classes corresponding to $n$, $\alpha(n)$, can be defined as
\begin{equation}\label{eq10}
 \alpha(n)=g(n)\tfrac{\ln n}{n},
\end{equation}
whose average $\langle \alpha(n)\rangle$ is equal to one.\\
This way, the contribution of periodic orbits to the Selberg trace formula can be restated as
\begin{equation}\label{eq11}
 d_{osc}(E)=\tfrac{2}{\pi k}\sum_{n=n_0}^{n=\infty}\alpha(n)\cos(2l\ln n).
\end{equation}
\\
In the semiclassical limit, the results obtained in the quantum regime exactly coincide with those obtained in the case of the first-order term of the WKB approximation for the wavefunction, i.e. when a 'constant-phase' approximation is performed. As a results, the energy levels of the eigenvalue problem for the Laplace-Beltrami operator are not modified, and the standard interpretation for the semiclassical limit of the wavefunction, i.e. the evaluation of the quantum wavefunction on the classical trajectory is enforced.\\
The geodesics that correspond to periodic orbits of the big billiard can be identified by a suitable subgroup of that defining the desymmetrized domain, such that the complete unquotiented geodesics can be reduced to a curve defined within the fundamental domain, and which consists of segments of geodesics.
\subsection{The Selberg trace formula for the modular billiard}
A specification of the Selberg trace formula for the modular billiard has been presented in \cite{bogomolny1}, for a different specification of the domain of the billiard. The main results are here summarized for comparison with the present approach, for which the shape of the billiard is not relevant.\\
\\
The trace formula for the modular billiard specifies then as
\begin{equation}\label{eq12}
d_{bil}(E)=\langle d(E)\rangle+\tilde{d}(E)+d_{osc}(E).
\end{equation}
Here, $\langle d(E)\rangle$ is the smooth part of the level density, and corresponds to $\langle d(E)\rangle=A/2\pi$, where $A$ is the area of the billiard domain.\\
Furthermore, the contribution due to periodic phenomena, i.e. $d_{osc}(E)$, reads
\begin{equation}\label{eq13}
 d_{osc}(E)=\sum_{n=n_0}^{n=\infty}\left( a^{(+)}(n)+\epsilon a^{(-)}(n)\right)\cos(2k\ln n),
\end{equation}
where $a^{(\pm)}(n)$ are the normalized multiplicities of periodic orbits corresponding to a subclass of conjugated matrices, whose determinants is $\pm1$, and are defined as
\begin{equation}\label{eq14}
 a^{(\pm)}(n)=2g^{\pm}(n)\tfrac{\ln n}{n}.
\end{equation}
Here, matrices with determinant $+1(-1)$ correspond to periodic orbits with an even (odd) number of reflections, and $g^{\pm}(n)$ is the number of periodic orbits with determinant $\pm1$ and trace (length) $n$. Furthermore, the prefactor $\epsilon$ accounts for the parity of the corresponding composition of matrices.\\
\paragraph{The assumption of Markov processes\label{mrkev}}
The modular group can be interpreted as generated by the two elements $s$ and $t$, i.e. a reflection $s$ and the $\sigma$-iterate of a translation $t$, with $\sigma\equiv\pm1$,
with $s^2=t^3=1$. This way, any elements of the modular group can be expressed as a composition of matrices whose elements are $s$ and $t^\sigma$, with $\sigma=\pm1$, and the expression of these composition of matrices are unique. To each conjugacy class, there is a corresponding composition of matrices starting with $s$ and ending with $t^\sigma$, up to cyclic permutations. It is therefore equivalent to consider composition of matrices constructed by the matrices
\begin{equation}\label{eq15}
 m_1=st\qquad m_2=st^{-1}.
\end{equation}
Within this framework, the sum can be performed over the probability for a conjugacy class  corresponding to a length $n$ to have a trace $r$ modulo $q$. The average over $n$ can be substituted by the average over all conjugacy classes for which $k$ symbols are requested; the two kinds of averages are claimed to be equivalent for sufficiently large values of $k$. Under this hypotheses, the composition of matrices of this scheme, which corresponds to the periodic orbits of the billiard, are generated as a Markov process, for which the probability for $m_1$ equals that for $m_2$, i.e. $1/2$. If the process is ergodic, every matrix $M_q$ can be build as a suitable composition of the blocks $m_1$ and $m_2$ (while, if the process is not ergodic, every matrix $M_q$ can be built by a suitable composition of the blocks $s$ and $t^\sigma$, but not $m_1$ and $m_2$). For cosmological billiards, anyhow, the dynamics can be demonstrated to be ergodic.\\
For Markov processes, the probability $P_k$ for one of the two matrices $m_1$ and $m_2$ to appear at a certain step $k$ of the composition of matrices is $1/2$, such that the probability for the matrix $m_i$ (with $i+1,2$) is given by
\begin{equation}\label{eq16}
 P_k(m_i)=\tfrac{1}{2}\left( P_{k-1}(m_im_1^{-1})+P_{k-1}(m_im_2^{-1})\right)
\end{equation}
This way, the probability that a composition of matrices is given by a matrix of trace $r$ and modulo $q$ equals the ratio of the number of matrices in $M_q$ with trace $r$, which defines, on its turn, the definition of $\alpha$ in (\ref{eq14}).
\section{The Selberg trace formula for cosmological billiards\label{section4}}
It is now possible to collect all the results achieved in the previous section and write down the Selberg trace formula for cosmological billiards.\\
For this, one needs to examine the objects involved in the Selberg trace formula and to find, for the trace formula, a suitable characterization for such objects.\\
The trace formula for cosmological billiards is stated by specifying the expression given by \cite{bogomolny1} for the modular billiard (\ref{eq12}) as
\begin{equation}
d_{cosm \ \ bil}(E)=\langle d(E)\rangle+\tilde{d}(E)+d_{cosm \ \ bil}^{osc}(E).
\end{equation}
The billiard systems are characterized by the periodicity conditions
\begin{subequations}\label{ttu}
\begin{align}
&\tt{T}\tt{u}\equiv \tt{u},\label{ttuu}\\
&\Psi(\tt{u})=\Psi(\tt{T}\tt{u})=\tt{T}\Psi(\tt{u}),\label{ttupsi}
\end{align}
\end{subequations}
where the generic condition for a periodic phenomenon is given by (\ref{ttuu}), in which the map $\tt{T}$ is a generic expression for any billiard map, and the variable $\tt{u}$ is any statistical variable for the billiard map, 
such that the expression of the energy levels characterizing $\psi$ in (\ref{ttupsi}) is attributed to the continued-fraction decomposition of the statistical variable.\\
According to the analysis in the previous Sections, the contribution due to periodic trajectories can be rewritten as a sum over the initial configurations which originate periodic trajectories, $u^+k$, as
\be\label{selgen}
d_{osc}E=\sum_{u^+_k}\varepsilon_{\tt{T}}(u_k)W^{\mu}(u^+_k)du^+_k\cosh L^{\tt{T}}(u_k).
\ee
The factors in each summand of the sum (\ref{selgen}) are defined according to the initial configurations $u^+_k$ which define a periodic trajectory, and according to the considered map $\tt{T}$ in the definition of periodicity, which, on its turn, is based on the kind of billiard considered, i.e. if the big billiard corresponding to the pure gravitational picture, of to the small billiard, corresponding to the presence of inhomogeneities and therefore to the consideration of the dominant symmetry walls, on the kind of symmetry quotienting for the description of the two billiard systems, and on the kind of phenomena which have to be characterized, i.e. according to the role attributed to the variable $u^-$ for the one-variable maps and for the two-variable maps.\\
Furthermore, it will be more convenient to spell out the elements of the Selberg trace formula starting from definitions on the UPHP and on the restricted phase space, according to which description allows for the best characterization of the dynamics of these billiard systems.\\
\\
The hyperbolic length $L^{\tt{T}}(u^+_k)$ of a closed geodesics depends on the trace $\mathcal{M}$, with $\mathcal{M}$ integer, of the composition of matrices that implies this periodic trajectory, as in Eq. (\ref{eq8}). Given a sequence of eras $k=(n_1, n_2, ..., n_k)$ which generates a periodic orbit in the quotiented billiards, such that $\sum_{j=1}^{j=k}n_j=m$, then $\mathcal{M}=m$ for the unquotiented billiards and $\mathcal{M}=pm$ for the quotiented billiards. Therefore,
\be
L^{\tt{T}}(u_k)\equiv L_k
\ee
for the quotiented dynamics, and
\be
L^{\tt{T}}(u_k)=pL_k
\ee
for the unquotiented dynamics, where $p$ is the order of the Kasner transformation in (\ref{mp}).\\
\\
The parity $\varepsilon$ of a closed geodesics depends on the periodic sequence $k$ for the quotiented billiards, as the number of reflections implied in the quotiented maps. For the unquotiented billiards, the parity of a closed geodesics depends, differently, on $mp$.\\
As a result,
\begin{itemize}
	\item for the quotiented big billiard, $\varepsilon^{\tt{T}}(u^+)\equiv \epsilon_k\equiv(-1)^k$, as defined on the UPHP;
	\item for the unquotiented big billiard, $\varepsilon{\tt{T}}(u^+)\equiv \epsilon_k^p\equiv(-1)^{mp}$, as defined on the restricted phase space;
	\item for the quotiented small billiard, $\varepsilon^{\tt{T}}(u^+)\equiv\epsilon^\mu_k$, where $\epsilon^{II}_k\equiv(-1)^k$ for $\mu=II$ and $\epsilon^{I}_k\equiv(-1)^{2k}\equiv1$ for $\mu=1$, as from the number of reflections on the UPHP;
	\item for the unquotiented small billiard on the UPHP, $\varepsilon^{\tt{T}}(u^+)\equiv\epsilon^{\mu \ \ p}_k$.
\end{itemize}
%%%%%%%%%%%%%%%%%%%%%%%%%%%%%%%%%%%%%%%%%%%%%%%%%%%%%%%%%5
The density of invariant measure for the initial values of the variable $u^+$ depends uniquely on $u^+$.\\
As a result, the Selberg trace formula will be given as a sum over the initial configurations.\\
For this, it is mathematically well-posed to express the differential spacings $dE$ of the energy levels as a sum over the suitable functions of the hyperbolic lengths of closed geodesics weighted by the differential of measure $Wdu$ of the invariant measure. The definition of this normalized invariant measure, given in \cite{leciannew}, for the Kasner quotiented era map, is usefully compared with the definition of analogue invariant measures for the full epoch map explained in \cite{Damour:2010sz}.\\
\\
The transition from a sum weighted by the densities of invariant measure $W(u^+_k)du^+_k$ to a sum weighted by BKL probabilities is due to the description of the stochastic limit of the dynamics, which is expressed by the stochastic limit of the densities of measure $W$. Within this limit, one can appreciate that a transition form a sum over initial configurations to a sum over periodic sequences and suitable symmetry operations among the elements of these sequences, which correspond to suitable symmetry operations n the generators of the reflections on the sides of the billiard table, which, on their turn, correspond to the implementation of the symmetries of the metric tensor as found from the solution to the Einstein filed equations in the asymptotic limit to the cosmological singularity, is due to the formal integration of the densities of measure $W8u^+$ aver the initial configurations, where the result of the integration is expressed in the limit to a stochastizing dynamics as a sum over the possible configurations, normalized by the average on the configurations considered equivalent under the limit to a stochastic process, within the role of the variable $u^-$, i.e. according to the descriptions obtained for the one-variable maps and for the two-variable maps.\\
\\
The BKL probabilities for sequences of eras depend on the exact sequence $k$.\\
For this, it is possible to express the Selberg trace formula by a sum over the integers $\mathcal{M}$, i.e the trace of the suitable composition of matrices, and on the suitable symmetry operations (and their elements in the the representatives of the suitable conjugacy subclasses of a matrix of trace $\mathcal{M}$) of sequences $k$ such that $\sum_j n_j=m$.  The specification of the spectral formula for the one-variable map or for the two-variable map implies a different characterization of the stochastic limit of the BKL probabilities for cosmological billiards, specified for different symmetry operations.\\
The order of the Kasner transformation which characterizes the unquotiented dynamics is determined by the features of a sequence $k$, such that $p\equiv p(k)$ for each sequence, and no sum over $p$ is implied.\\
As a result, the Selberg trace formula rewrites, in general, in terms of the BKL probabilities for the statistical maps, specified by the suitable symmetry operations on the generators of the iterations o fthe billaird maps, independently of the degree of stchastization of the dynamics, as a spectral distribution of the energy levels
\be\label{delta}
\tilde{\Delta} E_{cosm bill}=\sum_{\mathcal{M}}\sum_{per(k(\mathcal{M}))}\epsilon(k)P(k)_{symm}^{BKL}\cosh L_k,
\ee
where the equality implies on the rhs that the BKL probabilities evaluated for the closed geodesics are finite, in comparison with the differential version (\ref{selgen}). The suitable specifications for the symmetry-quoteiteng mechanisms are applied and commented in the next Subsections.
\\ 
\subsection{The Selberg trace formula for the quotiented big billiard}
It is now possible to rewrite contribution of periodic orbits to the Selberg trace formula for the quotiented big billiard. For this, one need to specify Eq. (\ref{eq11}) as a sum over the periodic initial configurations $u^+_{k}$ of the quotiented big billiard, where each summand contains the probability function density $W(u^+_k)du^+_k$, which, for the discrete periodic variables $u^+_k$ plays the role of a probability function density \cite{leciannew}, normalized according to the area of the big billiard table, and to explicitly define the prefactor $\epsilon(u^+_k)$ accounting for the numbers of reflections (i.e. for the determinant of the composition of matrices composing the iterations of the billiard map which fulfill the periodicity condition according to the quotiented big billiard era-transition map. For this, one
remarks that this prefactor is expressed by the number of eras defining the periodic configurations, i.e.
\be\label{epsilonk}
\epsilon{u^+_k}\equiv\epsilon(k)\equiv(-1)^{k}.
\ee
Accordingly, one obtains
\be\label{qbin}
d_{osc}(E)=\sum_{u^+_k}\epsilon_{u_k} W(u^+_k)du^+_k\cosh L(u^+_k),
\ee 
where $L(u^+_k)\equiv L_k$, i.e. the hyperbolic length of the closed geodesics depends only on the periodic set $k$ contained in the initial configuration $u^+_k$.
\\
This way, it is possible to specify the Selberg trace formula for the physical meaning of the one-variable map and for that of the two-variable map, that is, for the different statistical interpretation of the role of the variable $u^-$, by rewriting the sum over the initial configurations as a sum over the symmetry operations that characterize the two version of the billiard maps.\\
By applying Eq. (\ref{probcyc}) to the case of the two-variable map for the symmetry quotiented big billiard dynamics, i.e. by considering only the cyclic permutations among the components of each iteration of the CB-LKSKS map, the contribution to the Selberg trace formula for the two-variable big-billiard quotiented dynamics rewrites
\be\label{qb2}
\tilde{\Delta}_{osc}^{2-var}(E)=\sum_{\mathcal{M}}\sum_{per(k(\mathcal{M}))}\epsilon(k)P(k)_{cycl}^{BKL}\cosh L_k.
\ee 
On the contrary, in the case of the one-variable big billiard quotiented version, the sum has to be extended over all the permutations of the components of the billiard map which define the closed orbits, i.e. the application of Eq. (\ref{probper})
\be\label{qb1}
\tilde{\Delta}_{osc}^{1-var}(E)=\sum_\mathcal{M}\sum_{per(k)(\mathcal{M})}\epsilon(k)P^{BKL}_{per}(k) \cosh L_k.
\ee
This way, one sees that, differently from the definition of the modular billiard, the different symmetry operations among the generators of the group are considered only as mathematical tools that allow one to rearrange the iterations of the CB-LKSKS  billiard map \textit{without} modifying the umber of reflections contained in each summand, i.e. the physical interpretation of the billiard maps as encoding the statistical properties of the evolution of the scale factors is kept unmodified: the symmetry operations (cyclic permutations and exchange permutations) are considered only in the definition of the stochastic limit for the probability distributions, but do not directly define the hyperbolic length of the closed geodesics, even though these symmetry operations are generated by considering suitable commutators of the matrices composing the iterations of the CB-LKSKS map and would define 'unphysical bounces' and 'chopped segments of geodesics'.
\subsection{The Selberg trace formula for the quotiented small billiard}
It is now possible to write the Selberg trace formula for the small billiard, where the small billiard map on the UPHP is related to the big billiard maps by imposing the proper number of reflections which restore the equivalence of the dynamics n the two systems.\\
As a result, the sum over the initial configuration is split as
\be\label{qsin}
d_{osc}(E)=\sum_{u^+_k} \sum_{\mu}\epsilon^\mu_{u_k} W^\mu(u^+_k)du^+_k\cosh L(u^+_k),
\ee
where, in this case, the different number number of reflections which define the equivalence between the small billiard quotiented map and the big-billiard quotiented map does not modify the definition of the hyperbolic length of the closed periodic orbits $L(u^+_k)\equiv L_k$, which are straightforward defined in the case of the big billiard. On the contrary, the prefactors $\epsilon^\mu{u_k}$ are affected by the different number of reflections. The sum over $\mu$ is needed, as two different terms are obtained, i.e. those for the regions $I$ and $II$ of the restricted phase space. The prefactors $\epsilon^{\mu}_k$ write
\be
\label{epsilonkmu}
\epsilon^I_k\equiv\epsilon^{II}_k\equiv(-1)^{k+1},
\ee
where the last equivalence has been stated in comparison with Eq. (\ref{epsilonk}).\\
\\
Expressing the sum over the initial configurations as a sum over the symmetry operations that define the one-variable small billiard map and the two-variable small billiard map, under the limit of a stochastic proceed, requires to 'reshuffle' the sum over the different dynamical subregions of the restricted phase space by evaluating the proper probability, as for the stochastization of the dynamics, for a sequence to be described by points belonging to different dynamical subregions of the restricted phase space, independently of the symmetry operations required in the limiting process of the different maps. As a result, one learns that the the these limiting probabilities are expressed by the ratio of the area (according to the measure $\omega$) of the pertinent subregions of the restricted phase space with respect to the total area available for the dynamics (which corresponds to that available for the dynamics of the quotiented small billiard era maps). Of course, different specifications of the implications of the limit to a stochastic process can in principle provide with a slightly different expression of the sum over the symmetry operations, but do not modify the specification of the Selberg trace formula for the small billiard group and its expression for the statistical properties of the variable $u^-$, which can be interpreted as a further specification for the initial conditions in the one-variable map, while is attributed the specific task to transform the Gauss map for the continued-fraction decomposition of $u^+$ into a one-to-one mapping of the unit square onto the unit square by introducing the retrograde sequence of eras encoded in the fractional part of $u^-$, such that the fractional parts of the two statistical variables exhibit a stationary distribution.\\
Following these considerations, the contribution due to periodic trajectories to the Selberg trace formula for the two-variable small-billiard quotiented map rewrites
\be\label{qs2}
\tilde{\Delta}_{osc}^{2-var}(E)=\sum_{\mathcal{M}}\sum_{per(k(\mathcal{M}))}{\rm p}^{BKL}_{\mu cycl}(k)\epsilon(k)^\mu \cosh L_k,
\ee
where the definition of the prefactors $\epsilon^\mu_k$ has been specified according to Eq. (\ref{epsilonkmu}). Eq. (\ref{qs2}) is therefore the most direct characterization of the Selberg trace formula for the two-variable BKL map of the small billiard, where the presence of different reflections in the statistical map is encoded. For the characterization of the  Selberg trace formula as a sum over the suitable symmetry operations on the elements of the periodic sequence $k$, the presence of a different number of reflections is represented by the probability for a different number of reflections to be possible, independently of the number of epochs in each era of the map, as the ratio between the corresponding subregions $\mu$. As eras containing a smallish number of epochs are the most probable, it is straightforward to verify that this features is kept by the corresponding term in the Selberg trace formula.\\
\\
According to this specification, the relevance of the initial conditions for the variable $u^+$ is schematized, and the special role played by the eras with $n=1$ within the BKL statistics is outlined.\\
It is mandatory to observe that the two formulas admit a precise stochastic limit, but the exact behavior obtained for the sum over the initial trajectories is not recast, as the different BKL probabilities are summed with different prefactors. This is interpreted as an effect of the stochastization of the original BKL dynamics. This effect has to be compared with a different physical interpretation of the angular velocity at which the surface of section is crossed, as far as the full unprojected motion of the billiard ball is concerned.\\
\\
This way, for the two-variable map, suitable characterizations of the small-billiard map is found for the limit to a stochastic process according to the presence of a different number of reflections and according to the physical meaning of initial condition for the variable $u^+$. It is of interest to remark that it is possible to keep track only of the presence of a different number of reflections but not of the relevance of initial conditions, as in the first case with respect to the second one, but not the contrary.\\
\\
For the case of the one-variable small billiard map, considering Eq. (\ref{probper})  lead to the following expression
\be\label{qs1}
\tilde{\Delta}_{osc}^{1-var}(E)=\sum_{\mathcal{M}}\sum_{per(k(\mathcal{M}))}{\rm p}^{BKL}_{\mu per}(k)\epsilon(k)^\mu \cosh L_k,
\ee
Accordingly, the features of a stochastization of the dynamics, which, in the case of the big billiard, is expressed only by the independence of the probabilities on the order of the eras which compose the periodic trajectory and does not modify the definition of the length of the closed geodesics. On the contrary, the features of the small-billiard map are encoded on a specification of the stochastic probabilities for the different dynamical subregions of the restricted phase space.\\
\subsection{The  Selberg trace formula for the unquotiented big billiard}
The evaluation of the contribution due to the periodic trajectories in the case of the unquotiented small billiard map has to be followed by encoding in the sums the evaluation of the order of the order of the Kasner transformation in (\ref{mp}).\\
The application of this analysis allows one to outline tow main differences with respect to the quotiented maps.\\
On the one hand, the prefactor which encodes the sign of the determinant of the composition of the matrices which compose billiard map defining a periodic orbit will depend on the number of epochs in each era, according to the unquotiented big billiard map (\ref{bbz}), and not only on the number of Weyl reflections, i.e. one for each iteration of the quotiented map, as in the case of the quotiented big billiard.\\
On the other hand, the hyperbolic length of each periodic orbit will be a suitable multiple of the corresponding periodic orbit in the quotiented case, the multiplicity being expressed by the order of the Kasner transformation needed to fulfill Eq. (\ref{eq32}), but not on the number of reflections contained in the iterations of the billiard maps.\\
\\
As a result, the contribution due to periodic orbits for the Selberg trace formula is expressed by slitting the sum over the different orders of the Kasner transformations involved in the iterations of the quotiented billiard map, i.e.
\be\label{ubin}
d_{osc}(E)=\sum_{u^+_k}\epsilon^p_{u_k} W(u^+_k)du^+_k\cosh \left( p L(u^+_k)\right),
\ee
where the prefactor $\epsilon^p_{u_k}$ is defined as
\be
\epsilon^p_{u_k}\equiv\epsilon^p(k)\equiv(-1)^{mp}
\ee
through Eq. (\ref{mp}) and is not connected with the definitions of the same quantity for the quotiented dynamics. On the contrary, it depends on the periodic initial condition $u^+_k$ only according to the periodic configuration $k$.\\
As for the classification of the eigenvalues of the Laplace-Beltrami operator on the UPHP for the quotiented versions of the dynamics, the spectral formula depends on the initial value of the variable $u^+$ only as far as the invarianr density of measure for this countable set is concerned. On its turn, the initial values of the variable $u^+$ which define a periodic configuration in the quotiented version of the dynamics are able to determine the order of the Kasner transformation needed to recast Eq. (\ref{mp}) via the content of epochs of each era, and via the number of eras containing an even number of epochs. For each initial configuration $u^+$, therefore, a unique value of $p$ is implied, and no sum over $p$ is requested.\\
\\
As in the quotiented versions of the billiard maps, it is possible to evaluate the limit to a stochastic probability distribution by considering the different symmetry operations among the generators of the unquotiented billiard map, and by retaining the physical information contained in the different statistical maps.\\
Interestingly, the sum over the symmetry operations that define the two different maps is not affected by considering a different hyperbolic length of the periodic orbit corresponding to the same periodic configuration $k$: in fact, the sums over the two different symmetry operations among the generators of the unquotiented map takes into account only the physical trajectories, as the density of measure for the limiting process factors out. As one can straightforward verify by considering the different multiplicities implied for the BKL probabilities (\ref{probb}) for the stochastic limit, the multiplicities $p$ \textit{do} define the different probability compositions in the numerators and in the denominators of these definitions.\\
\\
For the case of the two-variable unquotiented billiard map Eq. (\ref{bbz}), the sum over the initial configurations is reconducted to a sum over the permutations of the generators of the two-variable map, 
\be\label{ub2}
\tilde{\Delta}^{osc}_{2-var}E=\sum_\mathcal{M}\sum_{per(pk(\mathcal{M}))}\left(P^{BKL}_{cycl}(pk)\epsilon^p_{u_k}\cosh pL_k\right).
\ee
\\
In the case of the one-variable unquotiented big billiard map, the specification of the pertinent contribution to the spectral formula as deriving from a permutation of the digits in the continued-fraction decomposition of the initial value of the variable $u^+$ that defines the periodic trajectory yields the expression
\be\label{ub1}
\tilde{\Delta}_{osc}^{1-var}E=\sum_{\mathcal{M}}\sum_{per(pk(\mathcal{M}))}\left(P^{BKL}_{per}(pk)\epsilon^p_{u_k}\cosh pL_k\right),
\ee
where one rewrites the sum over initial conditions as a sum over the pertinent permutations on the generators of the transformations considered for the unquotiented dynamics.\\
\\
In both cases, the sum over the initial values for the variable $u^+$ is 'redistributed' on a sum over the symmetry operations of the billiard maps, according to the physical interpretation of each conjugacy subclass, which define the dynamics only in the limit of a large number of iterations of the billiard maps.\\
Furthermore, in both cases, the substitution of the sum over the initial values of $u^+$ with a sum over the sequences $k$ is not able to attribute a specific value for $p$; this way, a sum over the three values of $p$ is requested. It is interesting to note that, in the case of the modular billiard, no such sum is present.\\
%%%%%%%%%%%%%%%%%%%%%%%%%%%%%%%%%%%%%%%%%
\subsection{The Selberg trace formula for the unquotiented small billiard}
The unquotiented small billiard map is obtained by applying the same machinery developed for the determination of the order of the Kasner transformation in the case of the big billiard to the different dynamical subregions of the restricted phase space, for which a different number of reflections is implied i the small billiard quotiented map. As a result, the sum over the initial values of the variable $u^+$ has to be further specified as
\be\label{usin}
d_{osc}(E)=\sum_{u^+_k}\sum_\mu  W^\mu(u^+_k)du^+_k\epsilon^{\mu\ \ p}_{u_k}\cosh \left( pL(u^+_k)\right),
\ee
where the prefactors $\epsilon^{\mu\ \ p}_{u_k}$ are here defined according to the different dynamical subregions of the restricted phase space where the corresponding density of invariant measure $W^\mu(u^+_k)du^+_k$ is evaluated, where $W^\mu(u^+_k)$ does not depend on the order $p$ of the Kasner transformation fulfilling Eq. (\ref{eq32}). 
The prefactors $\epsilon^{\mu\ \ p}$ are here defined as
\begin{subequations}
\begin{align}
&\epsilon^{I\ \ p}(u^+_k)\equiv \epsilon^{I\ \ p}(k)\equiv (-1)^{mp+1}\\
&\epsilon^{II\ \ p}(u^+_k)\equiv \epsilon^{II\ \ p}(k)\equiv (-1)^{mp}\equiv-\epsilon^{I\ \ p}(k),
\end{align}
\end{subequations}
such that they depend on both the order $p$ of the Kasner transformation on which a sum is performed and on the total number of epochs contained in the length $L_k$, according of the different prescriptions (\ref{epsilonkmu}), and the hyperbolic length of the periodic orbits is a multiple, according to the order $p$, of the length considered in the quotiented version of the dynamics.\\
\\
The sum over the symmetry operations which define the limit of a stochastic process of the dynamics and that allow one to express, within this limit, the sum over the periodic configurations to a sum over the elements of the suitable conjugacy subclasses need one to define a suitable modification of the coefficients for each function of the hyperbolic lengths $L_k$.\\
\\
In this case of the two-variable map, following the same reasoning adopted in the quotiented version, the sum over the cyclic permutations of the generators of the iterations of the unquotiented small billiard map implies a sum over the corresponding limit of the BKL probabilities for this symmetry operation
\be\label{us2}
\tilde{\Delta}_{osc}^{2-var}(E)=\sum_{\mathcal{M}}\sum_{per(pk(\mathcal{M}))}{\rm p}^{BKL}_{\mu cycl}(pk)\epsilon^{I\ \ p}_k \cosh p L_k.
\ee 
The Selberg trace formula for the unquotiented small billiard two-variable map outlines the relevance of the initial conditions for the solution to the Einstein field equations in the asymptotic limit towards the cosmological singularity, takes into account the presence of a different number of reflections implied by the statistical map, and selects the order $p$ of the Kasner transformation, which allows one to relate the periodicity phenomena of the quotiented big billiard and those of the unquotiented big billiard.\\
\\
The sum over all the permutations of the generators of the uqnquotiented small billiard maps for the BKL probabilities leads to the Selberg trace formula for the unquotiented small billiard described by the one-variable statistical map
\be\label{us1}
\tilde{\Delta}_{osc}^{1-var}(E)=\sum_{\mathcal{M}}\sum_{per(pk(\mathcal{M}))}{\rm p}^{BKL}_{\mu per}(k)\epsilon^{\mu\ \ p}_k \cosh p L_k.
\ee
For both the one-variable map and the two-variable map, the specification of the order of the Kasner transformations that allows for a definition of closed geodesics for the full unquotiented dynamics implies a sum over the order of the transformations for the different factors which define the probabilities for points on the periodic sequences to be issued from the different dynamics subregions of the restricted phase space.\\
\\
Once  more, one learn that the stochastic limit for the dynamics of cosmological billiards is understood in the different version of the Selberg trace formula as the definition of the probability for certain configurations to take place. In the case (\ref{stf}), the relevance of the initial conditions in the definition of the Selberg trace formula is therefore shifted form the probability for initial conditions to the definition of the parity of a periodic orbit. The relevance of the initial condition allows one to write, in the sum over the hyperbolic lengths of the closed geodesics, different summands, which admit a well-posed stochastic limits as far as the BKL probabilities are concerned, but for which the number of epochs contained in the first eras determine non trivial elements, such that the asymptotic stochastic limit, characterized by a factor $n_{per}(pk)$, does not simplify the numerators.\\
These features are not present in the definition of the stochastic limit for the big billiard dynamics, and are therefore considered as a new feature of the Selberg trace formula in the stochastized version of the dynamics, whatever the choice of the definition of the parity according tot he different dynamical subregions of the restricted phase space.
\subsection{The Selberg trace formula and the definition of boundary conditions}
In the quantum regime, the parity of the wavefucntions defines the boundary conditions. The parity is considered as the number of reflections contained in each composition of matrices corresponding to a given hyperbolic length of the pertinent closed geodesics.\\
For the modular billiard, the trace of such composition of matrices is shown to equal the hyperbolic sine (instead of the cosine used here) of the hyperbolic length of the periodic orbit in the case of lengths calculated from a composition of transformations of odd parity. Furthermore, the prefactor $\epsilon$ accounting for the parity of the wavefucntions is multiplied times an extra minus sign for the same cases.\\
Here, in the proposed decomposition of the sum corresponding to the features of the metric tensor, the hyperbolic length of a closed geodesics is computed always from the classical definition. Indeed, in the quantum version of the model, the role of the classical trajectories is enforced in the quantum regime via its semiclassical limit, their physical interpretation remains therefore the same.\\
\\
As in the previous literature, the parity of the quantum wavefucntions is determined by the number of reflections contained in the determinant of the considered composition of matrices.\\
In the symmetry-quotienting mechanisms of the dynamics, when the sum is extended over the initial values for the variable $U^+$, the parity of each contribution to the energy levels is specified not only according to the number of reflections, which corresponds to the number of bounces against the billiard walls, but also according to the reflection contained in the definition of the symmetry-quotiented maps. As an example, one sees that the parity of the energy levels associated to a closed geodesics is not accounted for by the total number of bounces (epochs), but only by the number of eras contained in the periodic sequence. Furthermore, in the case of the unquotiented big billiard dynamics, the parity is determined also by the order of the Kasner transformation, which is needed for the definition of periodicity in the unquotiented dynamics.\\
In the case of the small billiard maps, the definition of the parity becomes even more complicated. The different number of reflections corresponding to the billiard maps evaluated for different points in the restricted phase space is due to the need to restate the dynamics of the small billiard in a unique correspondence with that of the big billiard. This task in not achieved by the definition of maps in the restricted phase space only, where the definition of the implementation of a paradigm for iteration of the small billiard map in the restricted phase space appears of difficult practical use, as the subregions of the restricted phase space which determine the era map for the small billiard consists of curvilinear subdomains, and as the number of epochs in the small billiard differs form that defined for the big billiard for unpredictable 'steps'. If the billiard point of view is analyzed within the framework of the invariant quantities defined in classical Hamiltonian dynamics, one learns that the appearance of the extra reflection of some of the dynamical subregions of the restricted phase space is due to the fact that the Poincar\'e return map for the billiard ball in the case of the small billiard is not performed on a Poincar\'e surface of section corresponding to an entire gravitational wall: the definition of a symmetry quotienting mechanism in this case will take into account trajectories which are not defined according to the same direction of motion and orientations with respect to the billiard walls. Form the quantum point of view, therefore, the appearance of extra reflections in particular dynamical subregion of the restricted space space available for the quotiented dynamics of the small billiard is therefore interpreted as descending form the choice of a particular surface of section, and then by the need to relate the results with those obtained for the big billiard, where the Poincar\'e surface of section chosen for the Poincar\'e return map of the billiard ball corresponds to an entire gravitational wall.\\
These features are not removed when the symmetry quotienting mechanisms for the small billiard is' unfolded' and related with the full quotiented dynamics of the big billiard. In this case, in fact, the number of extra reflections is not canceled; on the contrary, these extra reflections 'daub' on the summands, according to the different orders of the Kasner transformations which recast the quotiented dynamics in the comparison with the big billiard. This phenomenon is  even more evident and more unexpected in the stochastic limits of the dynamics for the one-variable map and for the two-variable maps. In these cases, the stochastic properties acquired by the systems after a large number of iterations of the dynamics make the probability for these extra reflections to enter in the trace formula become the ratio between the corresponding subregions of the restricted phase space and the total area available for the dynamics: this ration corresponds to the probability for an epoch to be issued from a (previous with respect to the iterations of the billiard map) configuration where an extra reflection was contained in the billiard map. 
According to this discussion, one learns that, once more, the stochastic limit for the dynamics of cosmological billiards is understood as the corresponding limit of the probabilities for physical successions of bounces to take place. The presence of these extra reflections in the stochastic limit of the sum is therefore interpreted as inherited from the initial configuration for the definition of the parity of an orbit.
\subsection{Comparison with the description of a Markov process}
The specification of the Selberg trace formula for cosmological billiards here spelled out differs form the description of the same spectral formula reported in Paragraph \ref{mrkev} in terms of a Markov description of the dynamics for the different physical characterization of the billiard trajectories and for the different statistical properties attributed to them.\\
\\
Indeed, the description of a Markov limit for the dynamics is based on the assumption that the sequence of eras can be rewritten as a sequence of matrices implying a reflection and a translation. The symmetries of the Einstein field equations do not allow for such a schematization.\\
\\
The description in terms of a Markov process would furthermore describe the trajectories of cosmological billiards as consisting of disjoint segments of geodesics, where the continuous dynamics of cosmological billiards would be this way 'chopped'.\\
\\
The assumption of a Markovian characterization of the dynamics, for cosmological billiards, would moreover disregard the statistical implications of the BKL maps. Indeed, the statistical properties of cosmological billiard let one learn that one-epochs are the most frequent, while, fro a long sequence of epochs, a randomly-picked up epoch is mostly probable to belong to a long era. According to the Markov limit, any information about the BKL statistics would be overwritten by the equivalent $1/2$ Markov probabilities for the 'chopped' trajectories.\\ 
\\
Within the present work, differently, the dynamics of cosmological billiards is described by the BKL probabilities, which respect the BKL statistics and the symmetries of the Einstein field equations, and the connections between the initial conditions of the Einstein field equations and the different stages of the stochastization of the dynamics are analyzed, for a specification of the mathematical features of these quantities which is due to a classification of the initial conditions and to the determination for the countable set corresponding to periodic trajectories of suitable normalized probabilities.
%%%%%%%%%%%%%%%%%%%%%%%%%%%%%%%%%%%%%%%%%%%%%%%
\section{Scars in the wavefucntion of cosmological billiards\label{section5}}
From a mathematical point of view, the specifications of the Selberg trace formula for mathematical billiards is defined by considering only the number of conjugacy subclasses for a composition of matrices, which also corresponds to the class number of quadratic forms which define a quadratic equation for the endpoints $u^+$ and $u^-$ in the definition of a closed geodesics corresponding to a periodic orbits.\\
From a physical point of view, scars for the wavefucntion of cosmological billiards are accounted for the scars of the wavefuntion corresponding to the trajectories, which are the physical trajectories of these billiards, i.e. those accounted for the phenomena related to the statistical maps of cosmological billiards, for which a suitable expression of the Selberg trace formula is found by rewriting the sum over the initial conditions as a sum over the composition of matrices, which define the corresponding trajectories. Indeed, the $s$-fold symmetry of the solution to the Einstein field equations (\ref{abc}) picks out from all the possible billiard maps those which define the statistical maps, and {\it vice versa}.
\subsection{Scars in the wavefunction of the universe: numerical evidence}
The definition of BKL probabilities for the big billiard and for the small billiard, both in the symmetry-quotiented version and in the full unquotiented schematization of the discretized dynamics, both in the early-time BKL regime of the dynamics, as well as in the stochastizing BKL dynamics, through its steps, as well as the full stochastized regime of the BKL dynamics, allows one to uncover evidence for the presence of scars in the wavefucntion of the universe, within the classical regime as well as in the semiclassical limit, connected by the non-ambiguous definition of the BKL trajectories, as soon as, as in the present analysis, the Selberg trace formula is evaluated according to the physical interpretation of the statistical maps, i.e. for the meaning of the symmetry operations on the generators of the reflections in the iterations of the billiard maps, rather than on the degeneracy of the representatives of the conjugacy subclasses of each hyperbolic lengths of periodic orbits.\\
The numerical evidence for scars in the wavefucntion of cosmological billiards is here described by the different values acquired by the terms in the Selberg trace formula, for the expression of the BKL probabilities for the different statistical maps provides different values, with respect to the standard expression found for mathematical billiards. The most significant cases are listed in the following Tables, for the case of the simplest sequence $k=1, 2, 3$.\\
By the comparison of the numerical values of these probabilities, the description of the stochastization of the BKL dynamics allows one to compare the most probable configurations for the different physical characterization of the statistical maps 
%%%%%%%%%%%%%%%%%%%%%%%%%%%%%%%%%%%%%
\begin{table}
\begin{center}
\begin{tabular}{ | l | l |}

\hline
    
$ k=(n_1, n_2, n_3) $ & $ (\ln 2)P^{BKL}(k)$\\

  \hline
  \hline
  $1, 2, 3$ &  $0.004535155$ \\
  \hline 
   $1, 3, 2$ &  $0.004819285$ \\
  \hline
  $2, 1, 3$ &  $0.004773279$ \\
  \hline
  $2, 3, 1$ &  $0.004819285$ \\
  \hline
  $3, 1, 2$ &  $0.004773279$ \\
  \hline
  $3, 2, 1$ &  $0.004535155$ \\
  \hline

\end{tabular}
    %\label{tab:secondlevel}
\end{center}
\caption{\label{tableP} The values of the BKL probability $P^{BKL}(k)$(multiplied by the prefactor $\ln2$ to better appreciate the numerical values within the precision of the software) for the different permutations of the sequence $k=1, 2, 3$.}
\end{table}
%%%%%%%%%%%%%%%%%%%%%%%%%%%%%%%%%%%%%
\begin{table}
\begin{center}
\begin{tabular}{ | l | l |}

\hline
    
$ k=(n_1, n_2, n_3) $ & $ P^{BKL}_{1-var}(k)$\\

  \hline
  \hline
  $1, 2, 3$ &  $0.16055636$ \\
  \hline 
   $1, 3, 2$ &  $0.1705613270$ \\
  \hline
  $2, 1, 3$ &  $0.1689331094$ \\
  \hline
  $2, 3, 1$ &  $0.1705613270$ \\
  \hline
  $3, 1, 2$ &  $0.1689331094$ \\
  \hline
  $3, 2, 1$ &  $0.1605055636$ \\
  \hline

\end{tabular}
    %\label{tab:secondlevel}
\end{center}
\caption{\label{tableP1var} The values of the BKL probability $P^{BKL}_{1-var}(k)$ for the different permutations of the sequence $k=1, 2, 3$.}
\end{table}
%%%%%%%%%%%%%%%%%%%%%%%%%%%%%%%%%%%%%
\begin{table}
\begin{center}
\begin{tabular}{ | l | l |}

\hline
    
$ k=(n_1, n_2, n_3) $ & $ P^{BKL}_{2-var}(k)$\\

  \hline
  \hline
  $1, 2, 3$ &  $0.3210111271$ \\
  \hline 
   $2, 1, 3$ &  $0.3411226540$ \\
  \hline
  $3, 1, 2$ &  $0.3378662189$ \\
  \hline

\end{tabular}
    %\label{tab:secondlevel}
\end{center}
\caption{\label{tableP2var3} The values of the BKL probability $P^{BKL}_{2-var}(k)$ for the different permutations of the sequence $k=1, 2, 3$.}
\end{table}
%%%%%%%%%%%%%%%%%%%%%%%%%%%%%%%%%
\begin{table}
\begin{center}
\begin{tabular}{ | l | l |}

\hline
    
$ k=(n_1, n_3, n_2) $ & $ P^{BKL}_{2-var}(k)$\\

  \hline
  \hline
  $1, 3, 2$ &  $0.3411226540$ \\
  \hline 
   $2, 1, 3$ &  $0.3378662189$ \\
  \hline
  $3, 2, 1$ &  $0.3210111271$ \\
  \hline

\end{tabular}
    %\label{tab:secondlevel}
\end{center}
\caption{\label{tableP2var2} The values of the BKL probability $P^{BKL}_{2-var}(k)$ for the different permutations of the sequence $k=1, 3, 2$.}
\end{table}
%%%%%%%%%%%%%%%%%%%%%%%%%%%%%%%%%%%%
\begin{table}
\begin{center}
\begin{tabular}{ | l | l |}

\hline
    
$ k=(n_1, n_2, n_3) $ & $ 2 \ln 2\ \ {\rm p}^{BKL}_{I}(k)$\\

  \hline
  \hline
  $1, 2, 3$ &  $0.0009281969$ \\
  \hline 
   $1, 3, 2$ &  $0.0006829711$ \\
  \hline
  $2, 1, 3$ &  $0.002702109$ \\
  \hline
  $2, 3, 1$ &  $0.0022555405$ \\
  \hline
  $3, 1, 2$ &  $0.003249593$ \\
  \hline
  $3, 2, 1$ &  $0.002929456$ \\
  \hline

\end{tabular}
    %\label{tab:secondlevel}
\end{center}
\caption{\label{tablepI} The values of the BKL probability ${\rm p}^{BKL}_{I}(k)$ (multiplied by the prefactor $2\ln2$ to better appreciate the numerical values within the precision of the software) for the different permutations of the sequence $k=1, 2, 3$.}
\end{table}
%%%%%%%%%%%%%%%%%%%%%%%%%%%%%%%%%%%%%%%
\begin{table}
\begin{center}
\begin{tabular}{ | l | l |}

\hline
    
$ k=(n_1, n_2, n_3) $ & $ 2 \ln 2\ \ {\rm p}^{BKL}_{II}(k)$\\

  \hline
  \hline
  $1, 2, 3$ &  $0.003606959$ \\
  \hline 
  $1, 3, 2$ &  $0.004136314$ \\
  \hline
  $2, 1, 3$ &  $0.002071170$ \\
  \hline
  $2, 3, 1$ &  $0.002563746$ \\
  \hline
  $3, 1, 2$ &  $0.001523686$ \\
  \hline
  $3, 2, 1$ &  $0.001605700$ \\
  \hline

\end{tabular}
    %\label{tab:secondlevel}
\end{center}
\caption{\label{tablepII} The values of the BKL probability ${\rm p}^{BKL}_{II}(k)$ (multiplied by the prefactor $2\ln2$ to better appreciate the numerical values within the precision of the software) for the different permutations of the sequence $k=1, 2, 3$.}
\end{table}
%%%%%%%%%%%%%%%%%%%%%%%%%%%%%%%

\begin{table}
\begin{center}
\begin{tabular}{ | l | l |}

\hline
    
$ k=(n_1, n_2, n_3) $ & $ {\rm p}^{BKL}_{I\ \ 2-var}(k)$\\

  \hline
  \hline
  $1, 2, 3$ &  $0.1442793767$ \\
  \hline 
  $2, 3, 1$ &  $0.3506023101$ \\
  \hline
  $3, 1, 2$ &  $0.5051183132$ \\
  \hline

\end{tabular}
    %\label{tab:secondlevel}
\end{center}
\caption{\label{tablepI2var} The values of the BKL probability ${\rm p}^{BKL}_{I\ \ 2-var}(k)$ for the different permutations of the sequence $k=1, 2, 3$.}
\end{table}
%%%%%%%%%%%%%%%%%%%%%%%%%%%%%%%%%%%%%%%%%%%

\begin{table}
\begin{center}
\begin{tabular}{ | l | l |}

\hline
    
$ k=(n_1, n_2, n_3) $ & $ {\rm p}^{BKL}_{I\ \ 2-var}(k)$\\

  \hline
  \hline
  $1, 3, 2$ &  $0.1081585550$ \\
  \hline 
  $2, 1, 3$ &  $0.4279188459$ \\
  \hline
  $3, 2, 1$ &  $0.4639225992$ \\
  \hline

\end{tabular}
    %\label{tab:secondlevel}
\end{center}
\caption{\label{tablepI2var2} The values of the BKL probability ${\rm p}^{BKL}_{I\ \ 2-var}(k)$ for the different permutations of the sequence $k=1, 3, 2$.}
\end{table}
%%%%%%%%%%%%%%%%%%%%%%%%%%%%%%%%%
\begin{table}
\begin{center}
\begin{tabular}{ | l | l |}

\hline
    
$ k=(n_1, n_2, n_3) $ & $ {\rm p}^{BKL}_{II\ \ 2-var}(k)$\\

  \hline
  \hline
  $1, 2, 3$ &  $0.4687777109$ \\
  \hline 
  $2, 3, 1$ &  $0.3331967403$ \\
  \hline
  $3, 1, 2$ &  $0.1980255487$ \\
  \hline

\end{tabular}
    %\label{tab:secondlevel}
\end{center}
\caption{\label{tablepII2var} The values of the BKL probability ${\rm p}^{BKL}_{II\ \ 2-var}(k)$ for the different permutations of the sequence $k=1, 2, 3$.}
\end{table}
%%%%%%%%%%%%%%%%%%%%%%%%%%%%%%%%%%
\begin{table}
\begin{center}
\begin{tabular}{ | l | l |}

\hline
    
$ k=(n_1, n_2, n_3) $ & $ {\rm p}^{BKL}_{II\ \ 2-var}(k)$\\

  \hline
  \hline
  $2, 1, 3$ &  $0.2361322186$ \\
  \hline 
  $1, 3, 2$ &  $0.4715774184$ \\
  \hline
  $3, 2, 1$ &  $0.1830644049$ \\
  \hline

\end{tabular}
    %\label{tab:secondlevel}
\end{center}
\caption{\label{tablepII2var2} The values of the BKL probability ${\rm p}^{BKL}_{II\ \ 2-var}(k)$ for the different permutations of the sequence $k=1, 2, 3$.}
\end{table}
%%%%%%%%%%%%%%%%%%%%%%%%%%
\begin{table}
\begin{center}
\begin{tabular}{ | l | l |}

\hline
    
$ k=(n_1, n_2, n_3) $ & $ {\rm p}^{BKL}_{I\ \ 1-var}(k)$\\

  \hline
  \hline
  $1, 2, 3$ &  $0.07281194073$ \\
  \hline 
  $1, 3, 2$ &  $0.05357532572$ \\
  \hline
  $2, 1, 3$ &  $0.2119655866$ \\
  \hline
  $2, 3, 1$ &  $0.1769347443$ \\
  \hline
  $3, 1, 2$ &  $0.2549126946$ \\
  \hline
  $3, 2, 1$ &  $0.2297997081$ \\
  \hline

\end{tabular}
    %\label{tab:secondlevel}
\end{center}
\caption{\label{tablepI1var} The values of the BKL probability ${\rm p}^{BKL}_{I\ \ 1-var}(k)$ for the different permutations of the sequence $k=1, 2, 3$.}
\end{table}
%%%%%%%%%%%%%%%%%%%%%%%%%%%%%%%%%
\begin{table}
\begin{center}
\begin{tabular}{ | l | l |}

\hline
    
$ k=(n_1, n_2, n_3) $ & $ {\rm p}^{BKL}_{II\ \ 1-var}(k)$\\

  \hline
  \hline
  $1, 2, 3$ &  $0.2325933616$ \\
  \hline 
  $1, 3, 2$ &  $0.1335585996$ \\
  \hline
  $2, 1, 3$ &  $0.2667286149$ \\
  \hline
  $2, 3, 1$ &  $0.1653221732$ \\
  \hline
  $3, 1, 2$ &  $0.09825430475$ \\
  \hline
  $3, 2, 1$ &  $0.1035429459$ \\
  \hline

\end{tabular}
    %\label{tab:secondlevel}
\end{center}
\caption{\label{tablepII1var} The values of the BKL probability ${\rm p}^{BKL}_{II\ \ 1-var}(k)$ for the different permutations of the sequence $k=1, 2, 3$.}
\end{table}
%%%%%%%%%%%%%%%%%%%%%%%%%%%%%%%%%%
\begin{table}
\begin{center}
\begin{tabular}{ | l | l | l | l | l | l | }

\hline
    
$ k=(n_1, n_2, n_3) $ & $ P^S_{1-var}(k)$ & $ P^S_{2-var}(k)$ & $p$ & $P^S_{1-var}(pk)$ & $P^S_{2-var}(pk)$ \\

  \hline
  \hline
  $1, 2, 3$ &  $0.1\bar{6}$ & $0.\bar{3}$ & $2$ & $0.\bar{3}$ & $0.002...$  \\
  \hline

\end{tabular}
    %\label{tab:secondlevel}
\end{center}
\caption{\label{tablestochast} The values of the asymptotic limit towards the regime of complete stochastization of the BKL dynamics for the BKL probabilities for the one-variable map for the different permutations of the sequence $k=1, 2, 3$, for the two-variable maps for the cyclic permutations of the sequence $(1, 2, 3)$ and of the sequence $2, 1, 3$, in the quotiented version of the dynamics, and in its unquotiented version, for which $p=2$.}
\end{table}
%%%%%%%%%%%%%%%%%%%%%%%%%%%%%%%%%%%%%%%

\subsection{Scars in the wavefucntion of the universe: theoretical investigation}
The presence of scars in the wavefucntion is described as the presence of enhancements in the wavefunction in correspondence of the lowest-order period of closed geodesics for the quantum version of systems, which are classically described as chaotic.\\
\paragraph{Mathematical description of periodic trajectories in billiards}Furthermore, the presence of scars in the wavefucntions is connected with he appearance of the corresponding periodic orbits. The description of scars is therefore rendered even more complicated by the fact that the precise mechanism for which the periodic trajectories are originated has not been clarified yet form a mathematical point of view.\\
Indeed, the appearance of the periodic trajectories has been compared to the arithmetical nature of the groups which define the symmetries of the billiard tables in \cite{arith1} and \cite{arith2}, or it has been ascribed to the definition of the pattern with which the eigenvalues of matrices in random matrix theory \cite{matrix1}, \cite{matrix2} are implied. A comparison can be accomplished between the expression for the spacings between successive aigenvalued of the Selberg trace fromula for cosmological billiards here presented, both for the case of a sum over configurations for the exact densities of invariant distributions (\ref{selgen}), or in the case of a sum over the BKL probabilties (\ref{delta}), with the very recent investigation \cite{ata}, where distribution of the spacings (a suitable function of them) is analyzed in the setting of matrix theory, and compared with the zeroes of the Riemann $\zeta$ function, where tho the latter case should converge the spectral behavios of cosmological billiards as well. Furthermore, the different degrees of stochastization of the BKL probabilities within the evidence of scars in the wavefucntion of cosmological billiards, the random-matrix properties of the orbits, on which the wavefucntion is scarred, and the phase-space analysis of cosmological billairds can be compsred with the properties of the spectral formula when lax matrices are cocnerned \cite{lax}.\\
\\
The appearance of periodic continued fractions in number theory is not accounted for a suitable mathematical description, as the probability with which periodic continued fractions are generated, i.e. the expression of the probability for the digits of a purely periodic continued fraction to acquire specified values is not precisely described by the Gauss-Kuzmin theorem; the most modern approach to this problem has been proposed by indicating possible numerical simulations for these probabilities \cite{mnt}.\\
\\
\paragraph{Scars for the wavefunction of billiards}
Within the framework of cosmological billiards, periodic trajectories are originated by the statistical distribution of the initial conditions defines by the variable $u^+$ and $u^-$ for the Einstein field equations. The statistical distribution for these values is therefore implies by the features of the BKL statistical maps. A mathematical expression for the probabilities of orbits to happen within these billiards is given by the normalized $W(u^+_k)du^+_k$, which corresponds to the invariant density of measure for the iterations of the billiard maps, where the two tools are equiparated for discrete variables.\\
\\
It has been demosntrated that the big billiards is a suitable congruence subgroup of the small billiard, in any number of specetime dimensions, up to 11, such that the chaotic features of the system are conserved. Within this framework, a comparison is usefull, with an expansion of the Wel decomposition of the spectral distribution according to the group theoretical problem \cite{pavlov}.\\
\\
The statistical features of the modular billiard, described as in \cite{bogomolny1}, are based on the assumption of a Markov limit for the evolution of the dynamics, but also on the so-called Hardy-Littelwood method for the determination of the two-point correlation function, which is, on its turn, particularly appropriate for the description of arithmetical groups, as the degeneracy of the numbers of conjugacy subclasses of a matrix of trace smaller or equal to a certain integer value for arithmetical groups is the same as the probability to find prime numbers smaller or equal to the same certain value.\\
Nevertheless, within the approach followed in the present work, the sum over the conjugacy subclasses have been replaced by a sum over physically-relevant transformations of the generators, which acquire non trivial weights in the sum for cosmological billiards. Therefore, any overlapping of the present results with those obtained within the method followed in \cite{bogomolny1} is not of the strictest physical interpretation, as, in the expressions for the Selberg trace formula for cosmological billiards, the influence of the initial conditions on the stochastizing dynamics and the features of the statistical maps, which request the definition of new probabilities, the sum over the elements of the conjugacy subclasses do not average trivially.\\
\\
Within the framework of the dynamics of cosmological billiards, scars have analyzed according to different procedures.\\
The construction of Farey maps for cosmological billiards \cite{cornish} allows one to obtain a description of the orbits of these billiards, where periodic irrationals are isolated by devoiding the spacing among them, thus neglecting singular trajectories and non-periodic ones. In \cite{lecian}, the possibility to implement a phenomenologically-modified description of the restricted phase space for the implementation of such a map for cosmological billiards within the framework of cosmological billiards has been proposed. In particular, the comparison with Farey maps has been accomplished, differently form the procedure followed in \cite{cornish} by considering the BKL map as implemented on a phenomenological modification of the restricted phase space, for which the appearance of the simplest periodic trajectories in the first stages of the dynamics, i.e. for the first iterations of the billiard maps, in favored by a restriction of the range of the allowed values for the variable $u^+$ in the vicinity of the value corresponding to the lowest silver rations, which define the simplest periodic trajectories. In this model, which is aimed at a comparison with the implementation of a Farey map, only the effect of Kasner-quotiented maps are taken into account, i.e. the effects of the definition of periodicity (\ref{mp}) are not implemented.\\
\\
The results of the different expressions of the Selberg trace formula presented in the previous sections allows one to infer several items of information, which are useful in the analysis of the occurrence of particular structures for the lowest-order period closed geodesics.\\
Indeed, one notes that the expression for the one-variable map and for the two-variable map are characterized by the presence of non-trivial factors, which prevent the stochastic limit of the dynamics to imply a straightforward simplification of the asymptotic values of the denominators of the BKL probabilities defined for a stochastizing version of the dynamics. As a result, the lowest-order closed geodesics are accounted for in the expressions of the Selberg trace formula with smaller denominators, such that their weight is more relevant i the description of the evolution of the dynamics. The correspondence between the classical trajectory and the semiclassical limit, in which the quantum wavefunction is evaluated on the classical trajectory of cosmological billiards, has been discussed in \cite{lecian}. As the Selberg trace formula is valid independently of the physical regime analyzed, in the quantum model, the quantum wavefucntion is understood to express a higher probability for the system to be located in correspondence of the summands in the spectral formula, where the BKL probabilities are the highest. The presence of these non-trivial structures in the description of the energy level of the quantum model of cosmological billiards is therefore directly related with the enhancement of the absolute value of the wavefucntion in correspondence of the lowest- period closed geodesics, thus originating 'scars' in the wavefucntion.\\
From the analysis of \cite{levin2000}, the presence of scars in the wavefunction of cosmological billiards has been shown to be possibly related with the present observed large scale structure of the universe, according to the geometry describing the present universe.\\ 
\\
Within the present work, scars in the wavefucntion of the universe, as in the description of cosmological billiards, ar described by the accumulation of the wavefucntions on the classical periodic trajectories. The probability with which periodic trajectories are originated is described by the BKL statistics, where the normalized invariant density of measure for the variable $u^+$, $W(u^+)$, consists of the mass distribution function for discrete variables \cite{leciannew}.\\
The different weight for which the classically periodic trajectories are summed within the Selberg trace formula is due to the stochastization of the dynamics of cosmological billiard, which is implied by the iteration of the billiard maps.\\
The Selberg trace formula is initially given as a sum over the initial conditions; the stochastization of the dynamics allows one to rewrite this sum as one over the composition of matrices which define these trajectories; this phenomenon is accompanied by the different probabilities for which these trajectories are generated according to the physical characterization of the different statistical maps.\\
As a result, this analysis is found in agreement with the conjecture of Berry \cite{berri}, where scars in the wavefucntions of classically chaotic systems are due to the fact that the ergodicity properties of the classical motion allows one to consider that lowest-period periodic trajectories correspond to paths in the phase space, where the system is supposed to spend' a long time', also with respect to the Lyapunov stability of the model.\\
Moreover, the analysis of the classical Lyapunov instability of the billiard maps for the variable $u^+$ has to be compared, for cosmological billiards, with the possibility to consider the appearance of a quasi-isotropization mechanism, able to modify the strong anisotropic behavior of the model able to allow for a description of the complete thermal history of the universe.
\paragraph{The most recent characterizations}The presence of scars in the wavefucntion of classically chaotic systems has been related with different properties of these system, but not precise mathematical explanation has still been provided yet.\\
Very recently, new description of scars have been proposed, which account for different features of the wavefunction of classically chaotic systems, which had not been taken into account in the previous literature.\\
In \cite{fuchsian}, 
the presence of scars has been investigated for a broader class of models, i.e. those described by Fuchsian groups.\\
In \cite{nhurt}, the eigenvalues of the Laplace Beltrami operator have been connected with the length of periodic orbits, and the wavefucntion has been shown to permanently scar along some simplest periodic orbits. This analysis is based on the fact that the corresponding 'scarred' wavefucntions are shown to accumulate on the classically periodic trajectories.\\
In the new discussion of the phenomenon \cite{newchaos}. Some features of hyperbolic manifolds have been connected with those of arithmetical groups as far as scars are concerned.\\
\\
The presence of scars in cosmological billiards was first evidenced in \cite{levin2000} in the case of octagonal cosmological billiards, and in \cite{lecian}, for BKL cosmological billiards. For the simplest case of the periodic trajectories defined by the lowest silver ratios, in \cite{lecian}, the Farey map for the BKL mechanism defined in \cite{cornish} was implemented to the BKL statistic by a modification of the restricted phase space. In the present work, the restricted phase space is the one corresponding to the BKL statistics, and cases more general than the silver ratios have been examined in order to describe a greater variety f possible values acquired by the BKL probabilities, and for a numerical characterization of the stochastization process of the BKL dynamics defined in \cite{leciannew}.\\
\\
Within the present work, the most straightforward characterization of scars is that which finds agreement with the conjecture of \cite{berri}, i.e. that scars are obtained for the regions of the (restricted) phase space, where, within the classical regime, the billiard ball spends the longest time, i.e. where it is most probable to find the billiard ball according to ergodic properties of the system, such that the billiard ball is supposed to explore the complete phase space in the limit of an infinite time. The construction presented in this work leaves the phase space not modified, such that the conjecture of \cite{berri}, based on the ergodic properties of cosmological billiards, can be applied.\\
%%%%%%%%%%%%%%%%%%%%%%%%%%%%%%%%%%%%%%%%%%%%%%%%%
\section{The physical characterization of the model\label{section6}}
At the semiclassical limit, as well, a strong insight can be gained for the features of the transition form the Planckian era, whose energy scales are those which usually imply the presence of a quantum characterization of the gravitational interaction, to the classical regime, where the features of the spacetime are uniquely those described by the symmetries of the metric tensor and by the matter content in the Einstein field equations.\\
For this, it is possible to investigate the modifications to the energy levels caused by a quantum characterization of the gravitational interaction and those attributed to the properties of the wavefunction of the universe.\\
\\
As a result, and by taking into account features of the transition form the quantum regime of gravity to its classical description which are ensured by the validity of the Selberg trace formula, it is possible to relate the features observed for the geometrical properties of the present universe to those of the cosmological solutions studied for the asymptotic limit toward the cosmological singularity.\\
\\
From a more mathematical point of view, it is necessary to recall that the implementation of statistical maps for the description of the cosmological singularity is possible because it corresponds to encoding i these statistical maps the information contained in the group describing the small billiard domain and in its proper congruence subgroup describing the big billiard. For this, the characterization of the dynamics is substituted by the characterization of the symmetries of the congruence subgroup that described the pure gravitational billiard.\\    
\\
The formal definition of the geometrical tools needed to perform the analysis of cosmological billiards has to be further specified with respect to the information which is gathered form the analysis chaotic quantum systems. In fact, the definition of Poincar\'e surfaces of sections for different kinds of billiards allows one to study the properties of the billiard systems as from a discretized version of the dynamics regardless to the different boundary conditions which have to be imposed for the definition of a quantum wavefunctions. In this respect, the analysis of these surface of section is relevant in describing these of the particular  statistical map (int he case of cosmological billiards, a transformation related to the Gauss map) which is implemented in the discretization of the dynamics.\\
There are several descriptions of the mechanisms able to implement quantum maps for classically chaotic systems, as far as a precise physical characterization of the quantities represented by the maps are concerned. For this, it is interesting to remark that there has been no attempt to further specify these mechanisms for cosmological billiards for all the features of their dynamics, as any approximations which can be considered has to be compared, in principle, with the symmetries of the solutions to the Einstein field equations. Nevertheless, the mechanisms under investigations for different kinds of  geometries offer a great variety of possibilities to decide the perspective from which the problem should be faced.\\
\\
The definition of all the properties of cosmological singularity is far from being complete, while several efforts are performed within the different directions in which the physical investigation is conducted, and the features of the cosmological singularity, which can be traced in the observational evidences for the present structure of the universe motivate different research lines.\\
On the other hand, the very general discussion of the several mechanisms which modify the billiard dynamics in \cite{belqim} allow also for the consideration of extra-dimensional structures, as the outcome of compactification mechanisms. As already pointed out, the interest in cosmological billiards has been newly enhanced by the definition of higher-dimensional models, for which the schematization of the solution to the Einstein field equations in a target space endowed with Lorentzian metric is still that of a billiard system. It is taken therefore expected that these higher-dimensional models should admit a four-dimensional correspondent scenario, where the extra dimensions should be treated by means of dimensional reduction, i.e. after a suitable compactification mechanism, which should account for the non-direct observation of these dimensions. The influence of a compactification mechanisms has been hypothesized in \cite{belqim} for the simplest toy-model of a complete scheme consisting of five dimensions, and accounted for  a Brans-Dicke model within the four-dimensional analysis.
\subsection{A physical characterization of the quasi-isotropization mechanisms}
The decomposition of the values of the statistical variables which define the billiard trajectories within the BKL paradigm are strictly defined, in Hamiltonian systems, as they define the angular velocity at which the Poincar\'e surface of section is crossed, independently form the physical characterization of the system, according to the validity of the Selberg trace formula. Quantum BKL number have been defined by evaluating the quantum wavefunction of the universe, i.e. the solution to the WDW equation, in the WKB expansion of the semiclassical limit, on the classical BKL trajectory. Quantum BKL probabilities are therefore defined as the evaluation of the squared absolute value of the wavefucntion on the corresponding subregions of the restricted phase space. The semiclassical limit of the wavefucntion evaluated on a trajectory specified according to the BKL paradigm has been demonstrated to be characterized by the features needed to describe the quantum system and the semiclassical limit from the corresponding statistical point of view as far as classical BKL probabilities are concerned.\\
Within the present framework, it is possible to further specify these expressions for the definition of the probability for the wavefucntion of the universe to be characterized by a semiclassical limit, where a sequence $k$ of trajectories is considered.\\
The Selberg trace formula has been written form cosmological billiards, as a sum over the initial configurations of the solution to the Einstein field equations; the sum over these configurations has been reconducted to a sum over the BKL probabilities for a periodic sequence to take place, and is endowed of the information about the degree of stochasticity acquired by the system after a large number of iterations of the billiard maps, as the sum can be specified by the evaluation of the BKL probabilities according to the consideration of the exact BKL dynamics, a stochastizing dynamics of the sully-stochastized process.\\
The Selberg trace formula for billiard is evaluated, from a mathematical point of view, by the expression of the two-point correlation function characterizing the billiard system on the UPHP. The evaluation of the BKL probabilities for a statistical map, within a symmetry-quotienting mechanisms, therefore mathematically characterizes the two-point correlation function for cosmological billiards within the stochastization process of the dynamics under the evolution of the billiard map.\\
The experimental evidence of the observation of the pattern of the anisotropic sky, for several phenomena, can provide cosmological billiards with a physical characterization of the quasi-isotropization mechanisms, which can therefore be interpreted as 'freezing' the evolution of the BKL dynamics at a given time. More in particular, the range of the digits that constitute the 'frozen' sequence $k$ are estimated by the expression of the small anisotropy for a given gravitational observed effect by considering that BKL probabilities are a monotonically-decreasing function of these digits, and by considering also that the BKL statistics, as far as the two-variable map is concerned, imply that a randomly-chosen epoch in a long sequence of epochs is most likely belonging to an era containing a large number of epochs.\\
Furthermore, the degree of stochasticity of the BKL dynamics on which the quasi-isotropization mechanism has stated playing a predominant role in encoded in the evaluation of the two-point correlation function, which can best characterize the observational evidence for small anisotropies in the large scale-structure of the present universe. This equivalence follow from the equivalence between the BKL probabilities and the two-point correlation function for the Selberg trace formula.\\
According to the determination of the degree of stochasticity of the cosmological billiard, i.e. the pure BKL statistics, the stochastizing regime or the fully-stochastized system, at which the quasi-isotropization mechanism has been applied determines, on its turn, within the ranges of the estimation, the age of the universe at which the quasi-isotropization mechanism has started acting, by means of the parametrization of geodesics on the UPHP and the corresponding expression of the kasner coefficients in the solution to the Einstein filed equations in the asymptotic limit towards the cosmological singularity, (\ref{abc}).\\
According to the determination of the time interval between the cosmological singularity and the 'freezing' of the strong anisotropic regime of the BKL dynamics it is therefore possible to determine the age of the universe ar the corresponding time, i.e. if the phenomenon has stated acting during below the Planck age, at the Planck length or during the classicalized dynamics.\\
By matching all these items of information, it is therefore possible to determine if the anisotropy phenomena observed in the present Sky patterns are due to quantum-gravitational effects or to the hypothesis of non-quantum mechanisms. 
\paragraph{The Large scale structure of the universe}
As pointed out in \cite{levin2000}, it is possible to connect the presence of scars in the wavefunction of the universe at the Planckian age to the present observed large scale structure of the universe.\\
According to the recent accomplishments in Theoretical physics as well as in Experimental Physics, two main additional encouraging points should be stressed.\\
On the one hand, the definition of the Selberg trace formula allows one to extend the validity of the conjecture of Berry about the nature of scars also to the classicalized universe, by means of the interpretation of the semiclassical limit, individuated by the age at which the isotropic volume of the universe equals the Planck length, at which the wavefunction can be evaluated on the classical periodic trajectory defined by the BKL map, as the pertinent order of the WKB approximation coincides with the complete treatment in the quantum regime and with the classical differential spacing of the eigenvalues of the Laplace Beltrami operator given by the hyperbolic length of periodic trajectories as defined by the classical billiard maps, which are a suitable subset, and , in particular, a suitable composition os, the generators of the group, which defines the tessellation of the UPHP by means of the smallest desymmetrized domain possible.\\
This way, all the consideration developed within the quantum regime can be kept for the classical description. Moreover, any modification to the distribution of the eigenvalues of the Laplace Beltrami operator, which can be supposed to be modified by quantum features of the gravitational interaction, as well as by semiclassical features of the gravitational interaction, such that a classicalization process of these quantum effects, can be understood within the broad theoretical framework of a quasi-isotropization mechanism.\\
\\
As far as the observation of the actual large-scale structure of the universe is concerned, it is mandatory to recall that there exist several effects, which are usually analyzed in view of the investigation of possible anisotropic behavior of the present structure of the universe. These effects can be classified not only according to the different physical phenomena which they explain, but also according to the typical length scales, i.e. distances, which are considered for the observation of these effects.\\
Furthermore, there can be an even further description of the observed phenomena, according to the typical length scales at which the anisotropic features of the universe are studied. This last perspective in one that best fits the present work: indeed, the discovery of anisotropic patterns in the Sky analyzed can be connected to the degree of stochastization of the BKL paradigm, at which the quasi-isotropization mechanism has stated playing a predominant role, which can be further connected with the age of the universe at which this phenomenon happened, estimated by the correspondence between the length of the geodesics on the UPHP (invariant with respect to the physical characterization of the age of the Universe, i.e. if during the quantum stage, at the Planck scale or at the classicalized time) and the time evolution of the components of the metric tensor, which is, obviously, represented by the alternation of the sides of the billiards, where the schematization of the billiard ball is set.\\
Moreover, it should not be forgot that the description of several anisotropic patterns at different length scales can be related either to different external contribution to the Einstein field equations at different ages of the universe, as well as to different stages at which a single external contribution might have exerted different influences according to its own evolution, and also according to the mutual interaction with the gravitational filed. Among these contribution to the Einstein field equations, also the compactification of the other spatial dimensions \cite{Henneaux:2006gp} \cite{Henneaux:2008eg} can be considered.\\
The presence of matter and its interaction with the gravitational filed has been characterized also as a non-commutative geometry framework 
\cite{Connes:1996gi},
a case of non-commutative BKL cosmologies has been analyzed in 
\cite{Estrada:2012wd} 
and the case of Robertson-Walker line element in
\cite{Chamseddine:2011ix}.
 
\paragraph{Observations\label{discuss}}Furthermore, each effect of the gravitational field observed in the large scale structure of the universe as from the present experimental evidence is characterized by advantages and disadvantages, which mostly depend on the specific features analyzed. As an example, the Sunyaev-Zel'dovich effect \cite{sun1},
\cite{sun2}, \cite{sun3}, \cite{sun4} is relevant 
\cite{carlstrom}, \cite{plancksun} in describing the anisotropy in the distribution of cluster of galaxies. The stochastization of the original BKL dynamics can, within this framrwork, be considered as qualifying the dynamics of spacetime in the vicinity of the cosmological singularity as far as this psoposed analysis is concerned. Differently, the catalogue of galaxies represent a huge supply of data, from which the two-point correlation function for different length scales can be able to provide with a quantitative measure for the degree of anisotropy of the typical length scale, which can allow one to estimate the age of the universe at which that particular anisotropic pattern has been defined \cite{wen1} \cite{wen2}.
\section{Concluding remarks}
The Selberg trace formula has been specified for the periodic orbits of cosmological billiards, i.e. for the asymptotic limit of the asymptotic limit of the solution to the Einstein field equations towards the cosmological singularity, within the most general characterization of the symmetries of the metric tensor, under the BKL paradigm, in $4=3+1$ spacetime dimensions.\\
\\
The definition of the Selberg trace formula for cosmological billiards allows one to describe several aspects of the dynamics of the asymptotic limit towards the cosmological singularity of the Einstein filed equations for a generic cosmological solution under the BKL paradigm, for which such a limit is characterized by the spatial decoupling of space-time points, whose properties are defined only by ordinary time derivatives.\\
The Selberg trace formula allows one to relate the spectrum of the eigenvalues of the Laplace-Beltrami operators on the UPHP with the probabilities for which periodic orbits of the billiard take place, and is independent of the specification of the regime under which the motion takes place, i.e. it matches the geometrical properties of the UPHP, for which the quantum wavefunction is decomposed according a Fourier decomposition in the $u$ direction, for which the symmetry is given by the periodicity properties of a Fourier decomposition for circular trajectories, and 
a decomposition of the same wavefunctions according to the modified Bessel functions of the second kind, which describe the topology of the space, where the motion of a dynamical system as that of a point particle bouncing on the surface of the unit hyperboloid take place, as far as the full unprojected motion is concerned before the Hamiltonian constraint is taken into account, and for which a definition of periodicity is different form that of the projected version of the dynamics.\\
\\
The quantum version of the model can be analyzed according to two main different directions.\\
On the one hand, the features of the models here analyzed are compatible with the study of the structures, which are usually found for the quantum version of classically chaotic systems, which characterize the wavefucntions in correspondence of the classical periodic configurations, accounted for the simplest periodic trajectories.\\
\\
On the other hand, the quantum wavefucntion of the universe is relevant as its features encode those of the universe below the Planckian era, as well as the properties exhibited by spacetime, at the ages at which quantum deformations of the geometry and of the symmetries are predicted as the effect of quantum features of the gravitational interaction. The exact classification of the energy levels of the wavefunction allows one to discriminate, between those phenomena which are able to remove the chaotic features of classical early cosmology, those phenomena which are able to modify the chaotic dynamics with respect to the classical regime without modifying the intrinsic chaos of the system, and those phenomena which drastically imply the breakdown of the know properties of spacetime.\\
The definition of the Selberg trace formula for cosmological billiards in $4=3+1$ space-time dimensions is relevant not only in relating the geometrical properties of hyperbolic spaces with the physical description of cosmological billiards and in matching their classical description with the quantum version of the models in their semiclassical limit, where quantum wavefunctions are evaluated on the classical trajectories in the pertinent order of the WKB approximation, but are a powerful tool for the investigation of cosmological billiards in higher space-time dimensions. In fact, the phenomenon of cosmological billiards in $4=3+1$ space-time dimensions is show to match the proper gravitational limit of higher-order unification theories, where the definition of a Lorentzian Kac-moody algebra for the metric reproducing the evolution of the scale factors in the suitable target space allows one to select, among all time all higher-dimensional billiards, those for which a correct physical interpretation is obtained.\\
Within this framework, in higher-dimensional models, the boundaries of the cosmological billiards are obtained from the constraints of the equations of motion, which account for both the particular geometrical features of higher dimensional spaces as well as for the presence of different contribution (i.e. different kind of matter) int eh Einstein field equations.\\
Within this framework, the description of cosmological billiards is mostly performed by using th geometrical properties described by the corresponding algebras, which define the geometrical properties of  higher-dimensional hyperbolic spaces by means of the opportune algebraic structures. The quaternion algebra \cite{hejal} and the generalization of these algebras for the structures needed to analyzed the features of cosmological billiards in a higher number of spacetime dimensions has been discussed in \\
In this respect, a description of the Markov processes within the framework of a quaternion algebra has been achieved in \cite{bogomolny}. From the physical point of view, a quaternion algebra is at the basis of the description of cosmological billiards in $5=4+1$ spacetime dimensions, which reflects the properties of the corresponding (generalized) hyperbolic upper Poincar\'e half hyper-plane. Such a procedure allow one to generalize the techniques used for quaternions also for the algebraic structures and for the geometric models obtained in the geometrical description of the corresponding higher-dimensional unification theories.\\
\\
An expression for the Selberg trace formula able to contain all the countable set of initial conditions which originate periodic trajectories has been therefore established by considering the normalized densities of invariant measures as proper probability mass functions for the countable set of initial conditions, such that these densities of measure account for different weight on the sum over the hyperbolic lengths of the closed geodesics.\\
For the general features of the validity of the Selberg trace formula, at the quantum level, the quantum wavefunction for the universe is demonstrated to be most probably found at the energy levels corresponding to the values of the Kasner coefficients, which are most probably assumed for the statistical properties of the BKL parametrization of the Einstein field equations, such that scars in the wavefucntions are originated according to this mechanism. At the semiclassical level, the WKB expansion of the wavefucntion according to the classical trajectory appears to be statistically motivated for those values of the Kasner coefficients, for which the classical trajectory is most probably found, and correspond to the energy levels, at which the wavefucntion is enhanced because of the appearance of non-trivial elements in the sum over the normalized statistical probabilities. At the classical level, the eigenvalues of the Laplace-Beltrami operator are those that characterize the angular velocity at which the Poincar\'e surface of section id crossed.\\
The paper has been organized as follows.\\
In Section \ref{section2}, the main features of the solution to the Einstein field equations in the limit towards the cosmological singularity under the assumption of the BKL paradigm have been revised.\\
In Section \ref{section3}, the features of the Selberg trace formula have been recalled.\\
In Section \ref{section4}, the Selberg trace formula for cosmological billiards has been stated, and the differences with the expressions found in previous literature have been briefly described.\\
In Section \ref{section5}, the presence of scars for the wavefucntion of the universe has been analyzed for the features which can be inferred from the present analysis.\\
In Section \ref{section6}, a physical characterization of the mathematical features exposed in the previous Sections has been provided.
%%%%%%%%%%%%%%%%%%%%%%%%%%%%%%%
%%%%%%%%%%%%%%%%%%%%%%%%%%%%%%%%%%%%%%%%%%%%%%
\begin{figure*}[htbp]
\begin{center}
 \includegraphics[width=0.7\textwidth]{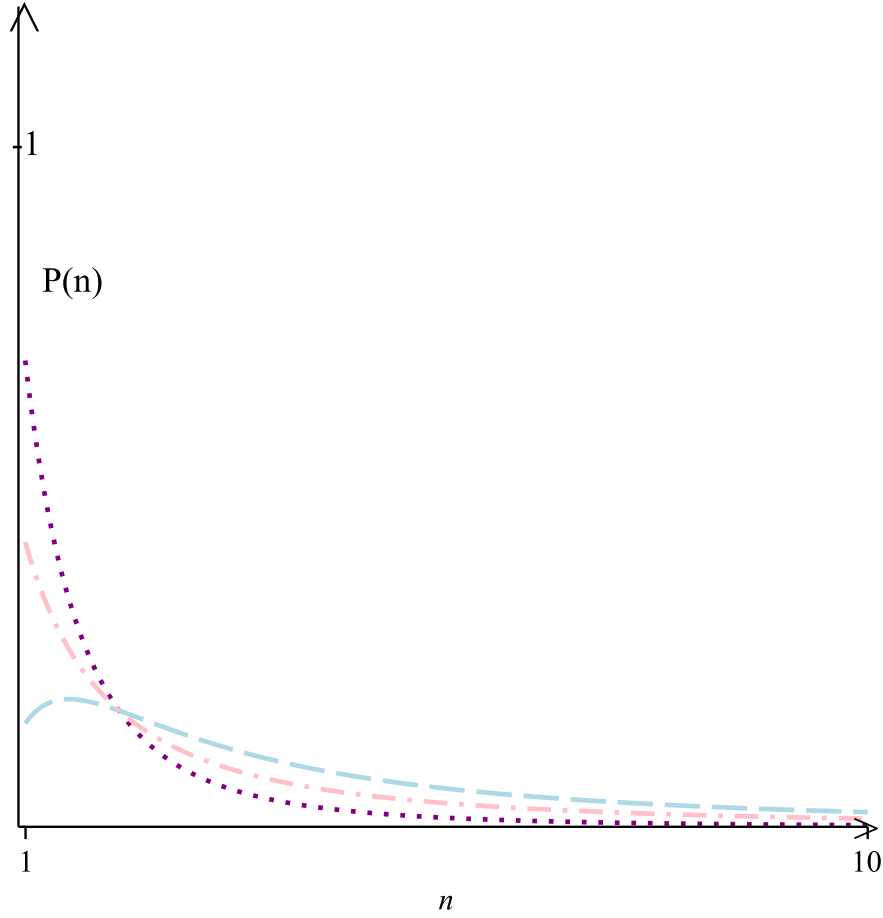}
\caption{\label{figura3} A comparison of the BKL probabilities for the big billiard and for the small billiard: the BKL probabilities $P^{BKL}(n)$ for the big billiard are plotted by the dashdot line; the BKL probabilities ${\rm p}^{BKL}_{I}(n)$ for the map (\ref{tsb2}) of the small billiard are plotted by a the dashed line; the BKL probabilities ${\rm p}^{BKL}_{II}(n)$ for the BKL map (\ref{tsb1}) are plotted by the dotted line. The particular features of the BKL probabilities ${\rm p}^{BKL}_{I}(n)$ for small $n$ are clarified, and the limit for large $n$ is illustrated.}
\end{center}
\end{figure*}
%%%%%%%%%%%%%%%%%%%%%%%%%%%%%%%%%%%%%%%%%%%%
\section*{Acknowledgments}
The work of OML was partially supported by the research grant 'Reflections on the Hyperbolic Plane' from the Albert Einstein Institute for Gravitational Physics - MPI, Potsdam-Golm, and partially by the research grant 'Classical and Quantum Physics of the Primordial Universe' from Sapienza University of Rome- Physics Department, Rome. OML acknowledges the Albert Einstein Institute for Gravitational Physics- MPI warmest hospitality during the corresponding stages of this work. OML is grateful to Prof. H. Nicolai for encouraging the investigation of periodic phenomena in cosmological billiards by outlining the relevance of Ref. \cite{Bogomolny:1992cj}, to Prof. P. De Bernardis for supporting this investigation, and to Prof. R. Ruffini for advising to investigate also the quantum features of the gravitational interaction. OML kindly thanks A. Kleinschmidt for the strictest discussion of these results and for the suggestion to implement a numerical investigation of the findings, and M. De Petris and G. Luzzi for encouraging discussions about \ref{discuss}. The environment Maple was used.\\
%%%%%%%%%%%%%%%%%%%%%%%%%%%%
%%%%%%%%%%%%%%%%%%%%%%%%%%%%%%%%%%%%%%%%%%%%%%%

\end{document}